\documentclass[11pt]{article}
\usepackage{pict2e}
\usepackage{amsmath,amssymb,graphicx}
\usepackage{hyperref}
\usepackage[nosort]{cite}
\usepackage[utf8]{inputenc} 
\usepackage[dvipsnames]{xcolor}
\usepackage{multicol,color,longtable}
\usepackage{array}
\newcolumntype{P}[1]{>{\centering\arraybackslash}p{#1}}
\usepackage{lscape,bbold}
\usepackage{multirow}

\definecolor{darkred}{rgb}{0.65,0.15,0}
\hypersetup{pdfborder={0 0 0},colorlinks=true,urlcolor=blue,citecolor=blue,linkcolor=darkred,linktocpage=true}

\newcommand{\nn}{\nonumber}

\newcommand{\fg}{{\mathfrak{g}}}
\newcommand{\so}{{\mathfrak{so}}}

\makeatletter

\@addtoreset{equation}{section}
\makeatother

\begin{document}

{\flushright  ICCUB-19-020\\[15mm]}

\begin{center}
{\LARGE \bf \sc  Newton--Hooke/Carrollian expansions of (A)dS\\[3mm] and Chern--Simons gravity}\\[5mm]

\vspace{6mm}
\normalsize
{\large  Joaquim Gomis${}^{1}$, Axel Kleinschmidt${}^{2,3}$, 
Jakob Palmkvist${}^{4}$, Patricio Salgado-Rebolledo${}^5$}

\vspace{10mm}
${}^1${\it Departament de F\'isica Qu\`antica i Astrof\'isica\\ and
Institut de Ci\`encies del Cosmos (ICCUB), Universitat de Barcelona\\ Mart\'i i Franqu\`es, ES-08028 Barcelona, Spain}
\vskip 1 em
${}^2${\it Max-Planck-Institut f\"{u}r Gravitationsphysik (Albert-Einstein-Institut)\\
Am M\"{u}hlenberg 1, DE-14476 Potsdam, Germany}
\vskip 1 em
${}^3${\it International Solvay Institutes\\
ULB-Campus Plaine CP231, BE-1050 Brussels, Belgium}
\vskip 1 em
${}^4${\it Department of Mathematical Sciences,
Chalmers University of Technology\\and University of Gothenburg, SE-412 96 G\"oteborg, Sweden}
\vskip 1 em
${}^5${\it Instituto de F\'isica, Pontificia Universidad Cat\'olica de Valpara\'iso\\ Casilla 4059, Valparaiso, Chile}
\vspace{15mm}

\hrule

\vspace{5mm}

 \begin{tabular}{p{14cm}}

We construct finite- and infinite-dimensional non-relativistic extensions of the Newton--Hooke and Carroll (A)dS algebras
using the algebra expansion method,
starting from the (anti-)de Sitter relativistic algebra in $D$ dimensions. These algebras are also shown to be embedded in different affine Kac--Moody algebras.
In the three-dimensional case, we construct Chern--Simons actions invariant under these symmetries. This leads to a sequence of non-relativistic gravity theories, 
where the simplest examples 
correspond to extended Newton--Hooke and extended \mbox{(post-)}Newtonian gravity together with their Carrollian counterparts.
\end{tabular}

\vspace{6mm}
\hrule
\end{center}

\newpage

\setcounter{tocdepth}{2}
\tableofcontents

\vspace{2mm}
\hrule

\section{Introduction}

Non-relativistic  symmetries are usually obtained from relativistic ones by means of Lie algebra contractions. The most well-known example 
is the Wigner--In\"on\"u contraction of the Poincar\'e algebra that leads to
the Galilei algebra~\cite{Inonu:1953sp} by sending the speed of light to infinity. Lie algebra contractions necessarily preserve the number of generators but can alter the cohomology of the Lie algebra, thus allowing for different central extensions like the Bargmann extension of the Galilei algebra~\cite{Bargmann:1954gh}, which is crucial when taking the limit of the relativistic particle action to the non-relativistic one~\cite{levy1969group,Marmo:1987rv}. 

In recent years, more general constructions of non-relativistic symmetry algebras from relativistic ones
have been explored. One example is given by the method of Lie algebra expansions~\cite{Hatsuda:2001pp,deAzcarraga:2002xi,izaurieta:2006zz} applied to the Poincar\'e algebra~\cite{Bergshoeff:2019ctr}. This method provides an infinite sequence of non-relativistic algebras extending the Galilei algebra with an increasing number of generators, which have been
used in~\cite{Hansen:2018ofj,Hansen:2019vqf,Bergshoeff:2019ctr,Ozdemir:2019orp} to construct various gravitational actions. The Lie algebra expansion method can also be related to a sequence of post-Newtonian limits as shown in~\cite{Gomis:2019sqv}, and has also been applied to derive diverse non-relativistic symmetries in the context of (super-)gravity~\cite{Gonzalez:2016xwo,Penafiel:2016ufo,deAzcarraga:2019mdn,Penafiel:2019czp,Concha:2019lhn}. Another method is based on a Galilean free Lie algebra~\cite{Gomis:2019fdh} that can be thought as the most general extension of the Galilei algebra and, upon taking quotients, has a connection to Lie algebra expansions and Kac--Moody algebras. One interesting conceptual point made in~\cite{Gomis:2019sqv} is that the sequence of Lie algebras naturally comes with a generalisation of Minkowski space with more coordinates and an extension of the Minkowski metric. Using this generalised Minkowski space it is possible to define particle actions invariant under these extended algebras that naturally incorporate post-Newtonian corrections in the non-relativistic limit.

In the present article, we study the non-relativistic symmetries obtained by Lie algebra expansion of the AdS or dS algebra in $D$ space-time dimensions, i.e.  $\mathfrak{so}(D-1,2)$ or $\mathfrak{so}(D,1)$. This generalises previous constructions to include a cosmological constant and generates an infinite family of algebras of Newton--Hooke \cite{Bacry:1968zf,derome1972hooke}
type. By taking the limit of the cosmological constant to zero one recovers the non-relativistic algebras introduced in~\cite{Hansen:2019vqf} and further studied in~\cite{Bergshoeff:2019ctr,Gomis:2019fdh,Gomis:2019sqv}. 

Besides the non-relativistic limit related to Galilean symmetries ($c\to\infty$) we also consider the case of ultra-relativistic Carrollian limit ($c\to 0$).\footnote{Work on Carroll symmetries in the context of electrodynamics and brane dynamics can be found for instance in~\cite{Duval:2014uoa,Clark:2016qbj,Basu:2018dub}.} As is known from~\cite{Barducci:2018wuj} the Carroll algebra can be understood by very specific changes in the commutation relations and associated changes in the starting point of the algebra expansion procedure. Applying the expansion procedure then produces an infinite family of Lie algebras associated with the Carrollian limit.
We shall also show how our construction in both cases is related to affine Kac--Moody algebras, where the role of the expansion parameter is taken by the spectral (or loop) parameter of the affine algebra. 

After the introduction of the infinite family of algebras we consider gravitational models based on them by focussing on the case of $2+1$ space-time dimensions. We construct Chern--Simons theories based on these algebras since they admit non-degenerate bilinear forms. We show that the family of algebras systematically generates non-relativistic gravity theories extended by a cosmological constant, such as extended Newton--Hooke gravity \cite{Papageorgiou:2010ud,Hartong:2016yrf}, that reduces to extended Bargmann gravity for $\Lambda\to 0$~\cite{Papageorgiou:2009zc,Bergshoeff:2016lwr}, and post-Newtonian gravity~\cite{Ozdemir:2019orp}\footnote{In~\cite{Ozdemir:2019orp} this was called `extended gravity' but in the light of the results of~\cite{Gomis:2019sqv} this is better understood as a post-Newtonian correction.}. 

Our construction also connects to recent discussions of non-relativistic expansions of the metric as described in~\cite{dautcourt1990newtonian,DePietri:1994je,Dautcourt:1996pm,VandenBleeken:2017rij,Hansen:2018ofj,Hansen:2019vqf,Hansen:2019svu}. This can be made very precise in the context of the Chern--Simons formulation and we shall elaborate on this connection in the conclusions.

In an appendix, we generalise the expansion procedure to obtain an infinite family of extensions of the non-relativistic Maxwell algebra. Moreover, we outline that our method is not only applicable non-relativistic symmetries corresponding to point particle limits, but also to extended objects where the $D$ covariant dimensions are split into $p+1$ and $D-p-1$ directions for a 
$p$-brane~\cite{Gomis:2005pg,Brugues:2006yd}. Similar constructions in the case without a cosmological constant have been considered previously in~\cite{Batlle:2016iel,Barducci:2018wuj,Bergshoeff:2019ctr,Harmark:2019upf}. 

\medskip

{\bf Note added.} While this manuscript was being finalised, the preprints \cite{Concha:2019dqs,Ali:2019jjp} appeared on the arXiv that have some overlap with some of our results.

\section{Expansions of the (A)dS algebra}

In this section, we will construct an infinite family of non-relativistic algebras of the Newton--Hooke and Carroll AdS type. The first members of this family will be given by the Newton--Hooke algebra in the former case and the Carroll AdS algebra in the latter (without extensions) 
\cite{Bacry:1968zf,derome1972hooke}. We also obtain an infinite-dimensional algebra that contains these and other intermediate cases, 
generalises the one obtained in \cite{Hansen:2019vqf,Bergshoeff:2019ctr,Gomis:2019fdh}, as quotients. All our algebras contain a cosmological constant as we start from the AdS algebra.

We will construct the series of non-relativistic algebras by means of Lie algebra expansions
\cite{Hatsuda:2001pp,deAzcarraga:2002xi,izaurieta:2006zz}. More precisely, we will use the semigroup expansion technique \cite{izaurieta:2006zz} with semigroup $S_{E}^{(N)}$, which will be defined below.

Our starting point is the (A)dS algebra in $D$ dimensions:
\begin{subequations}\label{AdSalgCommutators}
\begin{align}
[\tilde J_{AB}, \tilde P_{C}] & = 2\eta_{C[B} \tilde P_{A]}\, , \\
[\tilde J_{AB}, \tilde J_{CD}] &= 4\eta_{[A[C }\tilde J_{D]B]}\, ,\\
[\tilde P_{A}, \tilde P_{B}]& = -\Lambda \tilde J_{AB}\, ,
\end{align}
\end{subequations}
which denotes in a unified manner $\mathfrak{so}(D-1,2)$ (for $\Lambda<0$) and $\mathfrak{so}(D,1)$ (for $\Lambda>0$). 
Here, $\tilde P_{A}$ and $\tilde J_{AB}$ are the generators of spacetime translations and Lorentz transformations, respectively.  Capital indices run over $A=0,\ldots,D-1$ and the
Minkowski metric has been chosen to have mostly plus signature. 

In order to perform a non-relativistic expansion of the (A)dS algebra,  it is convenient to decompose the relativistic indices in the time and space components, 
$A=(0,a)$, where $a=1,\ldots,D-1$, and relabel the Lie algebra 
generators\footnote{Notice this convention is different from the one in 
\cite{Gomis:2019sqv} where $\tilde J_{a0}\equiv \tilde G_{a}$ was used.}
as 
\begin{align}\label{gendecpoincare}
\tilde J_{AB}&\rightarrow\{\tilde J_{0a}\equiv \tilde G_{a} \,,\;\tilde J_{ab}\}\, ,& 
\tilde P_{A}&\rightarrow\{ \tilde P_0\equiv \tilde H \,,\; \tilde P_{a}\}\, .
\end{align}
Then, the commutation relations \eqref{AdSalgCommutators} can be rewritten in the form
\begin{multicols}{2} 
\begin{subequations}\label{splitadsalg}
\allowdisplaybreaks
\setlength{\abovedisplayskip}{-15pt}
\begin{align}
[\tilde G_{a},\tilde H]&=\tilde P_{a}\, ,\\
[\tilde G_{a},\tilde P_{b}]&=\delta_{ab}\tilde H\, , \\
[\tilde J_{ab},\tilde P_{c}]&= 2\delta_{c[b}\tilde P_{a]}\, ,\\
[\tilde G_{a},\tilde G_{b}]&=\tilde J_{ab}\, ,\\
[\tilde J_{ab},\tilde G_{c}]&=2\delta_{c[b}\tilde G_{a]}\, ,\\
[\tilde J_{ab},\tilde J_{cd}]&=4\delta_{[a[c}\tilde J_{d]b]}\, ,\\
[\tilde P_{a},\tilde H]&=\Lambda \tilde G_{a}\, ,\\
[\tilde P_{a},\tilde P_{b}]&=-\Lambda \tilde J_{ab}\, .
\end{align}
\end{subequations}
\end{multicols}
\noindent We note that the decomposition~\eqref{gendecpoincare} is adapted to point particles in the sense that one direction---that can be thought of as the world-line direction of the particle---is singled out. In appendix~\ref{sec:brane}, we also consider the case of extended objects.

\subsection{Extended Newton--Hooke algebras}

In order to perform expansions of (A)dS of Newton--Hooke type, we first note that \eqref{splitadsalg} allows for the following subspace decomposition $V=V_{0}\oplus V_{1}$ \cite{Bergshoeff:2019ctr}
\begin{align}
V_{0}&=\{\tilde H, \tilde J_{ab}\}\,,& 
V_{1}&=\{\tilde P_{a},\tilde G_{a}\}\,,\label{splitAdS}
\end{align}
which satisfies a $\mathbb{Z}_{2}$-graded structure, i.e.,
\begin{align}\label{Z2Vs}
[V_{0},V_{0}]&\subset V_{0}\,,&  [V_{0},V_{1}]&\subset V_{1}\,,&  [V_{1},V_{1}]&\subset V_{0} \,.
\end{align}
This decomposition is similar to a symmetric space decomposition. The homogeneous coset space in the case of AdS is $SO(D-1,2)/(SO(D-1)\times \mathbb{R})$ and $SO(D,1)/(SO(D-1)\times \mathbb{R})$ in the dS case.

We will consider expansions with the semigroup $S_{E}^{(N)}=\{\lambda_{0},\dots,\lambda_{N+1}\}$, whose multiplication law is given by \cite{izaurieta:2006zz}
 \begin{equation}\label{semigroupsn}
\lambda_i \cdot\lambda_j =
\left\{ \begin{array}{ccc}
\lambda_{i+j} & {\rm if} & i+j\leq N\,,\\
\lambda_{N+1} & {\rm if} & i+j > N\,,
\end{array}\right.
 \end{equation}
where $\lambda_{N+1}$ acts as the zero of the semigroup, as it satisfies $\lambda_{N+1} \cdot \lambda_i =\lambda_i \cdot \lambda_{N+1}=\lambda_{N+1}$ for all $\lambda_i$. This semigroup
admits the subset decomposition
\begin{equation}\label{s0ands1}
\begin{aligned}
S_{0}^{(N)}&=s_{0}^{(N)}\cup\left\{ \lambda_{N+1}\right\} \,,& s_{0}^{(N)}&=\left\{ \lambda_{2m}\;|\;m=0,\dots,\left[\tfrac{N}{2}\right]\right\}\,,\\
S_{1}^{(N)}&=s_{1}^{(N)}\cup\left\{ \lambda_{N+1}\right\} \,,& s_{1}^{(N)}&=\left\{ \lambda_{2m+1}\;|\;m=0,\dots,\left[\tfrac{N-1}{2}\right]\right\}\,,
\end{aligned}
\end{equation}
which is is compatible with \eqref{Z2Vs} in the sense that
\begin{align}
S_{0}^{(N)}\cdot S_{0}^{(N)}&\subset S_{0}^{(N)},& S_{0}^{(N)}\cdot S_{1}^{(N)}&\subset S_{1}^{(N)}\,, & S_{1}^{(N)}\cdot S_{1}^{(N)}&\subset S_{0}^{(N)} \,,
\label{resdecsemigroup}
\end{align}
and therefore \emph{resonant} with the choice of $V_{0}$ and $V_{1}$ in \eqref{splitAdS}. 

Thus, a reduced resonant expanded algebra can be defined as the direct sum
 \begin{equation}\label{redresp}
\Big\{s_{0}^{(N)}\otimes V_0\Big\} \oplus
\Big\{s_{1}^{(N)}\otimes V_1 \Big\}
 \end{equation}
where the reduction condition in the algebra is implemented by the constraints
\begin{align}
\lambda_{N+1}\otimes \tilde J_{AB}&= 0\,,&\lambda_{N+1}\otimes \tilde P_A &=0\,,\label{redcon}
\end{align}
which map the zero $\lambda_{N+1}$ of the semigroup to the zero element in the expanded algebra. In the following we will show how, for different choices of the semigroup $S_{E}^{(N)}$, 
 the reduction \eqref{redresp} lead to a non-relativistic algebras that generalise the Newton--Hooke symmetry.

 The simplest example in our construction corresponds to the  expansion with the semigroup $S_{E}^{\left(1\right)}$, which  as we will see is equivalent
to a non-relativistic contraction of the (A)dS algebra. In other words, this expansion gives the Newton--Hooke algebras. By setting $N=1$ in \eqref{semigroupsn} and \eqref{redresp}, the generators of the expanded algebra are given by
\begin{equation}
\begin{aligned}
   J_{ab}&=\lambda_{0}\otimes \tilde  J_{ab}\,, &    G_{a}&=\lambda_{1}\otimes \tilde 
G_{a}\,,\\ 
H&=\lambda_{0}\otimes \tilde  H\,, &   P_a&=\lambda_{1}\otimes \tilde  P_{a}\,.\label{expgens1}
\end{aligned}
\end{equation}
Using the relativistic commutation relations \eqref{splitadsalg} together with the reduction condition \eqref{redcon} for the zero element $\lambda_2$, the commutation relations for the generators \eqref{expgens1} read
\begin{multicols}{2}
\begin{subequations}\label{galileanalgD}
\setlength{\abovedisplayskip}{-15pt}
\allowdisplaybreaks
\begin{align}
[  G_{a},  H]&=P_{a}\, ,\\
[  J_{ab},  P_{c}]&= 2\delta_{c[b}  P_{a]}\, ,\\
[  J_{ab},  G_{c}]&=2\delta_{c[b}  G_{a]}\, ,\\
[  J_{ab},  J_{cd}]&=4\delta_{[a[c}  J_{d]b]}\, ,\\
[  P_{a},   H]&=\Lambda   G_{a}\, ,
\end{align}
\end{subequations}
\end{multicols}\noindent
which corresponds to the Newton--Hooke$_{\mp}$ algebra (with $\Lambda\,\raisebox{.3\height}{\scalebox{.8}{$\lessgtr$}}\, 0$) without extensions \cite{Bacry:1968zf,derome1972hooke}, 
this algebra reduces to the Galilei symmetry for $\Lambda=0$.
Therefore the use of the semigroup $S_{E}^{\left(N\right)}$ will create a 
family of generalised Newton--Hooke algebras for all the possible values of $N$.

The case with $N=2$  leads to a Newton--Hooke algebra with some extensions. Unlike the previous case, when using the semigroup $S_{E}^{\left(2\right)}$, the element $\lambda_2$ is no longer the zero of the semigroup. In this case the reduction condition \eqref{redcon} holds for a new semigroup zero element $\lambda_3$, and \eqref{expgens1} has to be supplemented with two extra expanded generators
\begin{equation}
 S_{ab}=\lambda_{2}\otimes \tilde  J_{ab} \,,\hspace{10mm}
M=\lambda_{2}\otimes \tilde  H\,. \label{expgens2}
\end{equation}
Using the semigroup product law \eqref{semigroupsn} for $N=2$, we find that the non-vanishing commutation relations for the expanded algebra are given by the Newton--Hooke commutators \eqref{galileanalgD} together with
\begin{multicols}{2}
\begin{subequations}\label{ExtensionGalD}
\setlength{\abovedisplayskip}{-15pt}
\allowdisplaybreaks
\begin{align}
[  G_{a},  P_{b}]&=\delta_{ab}M\, , \\
[  G_{a},  G_{b}]&=S_{ab}\, ,\\
[  J_{ab},  S_{cd}]&=4\delta_{[a[c}  S_{d]b]}\,,\\
[  P_{a},   P_{b}]&=-\Lambda   S_{ab}\, .
\end{align}
\end{subequations}
\end{multicols}\noindent
This expansion produces the Bargmann central extension $M$ and a non-central extension $S_{ab}$ \cite{Bonanos:2008kr}. This algebra can be understood as the generalisation of the double central extension 
 \cite{medina1985algebres,Figueroa-OFarrill:1995opp,Matulich:2019cdo}
of the Newton--Hooke algebra to $D>3$. In fact, in 2+1 dimensions the generator $S_{ab}$ can be dualised to a scalar and becomes central. In that case the algebra \eqref{ExtensionGalD} becomes exactly the extended Newton--Hooke algebra ($\Lambda\neq 0$) \cite{arratia1999classical,Alvarez:2007fw} or the extended Bargmann algebra ($\Lambda=0$) \cite{levy1971galilei,Brihaye:1995nv}.

\subsubsection*{Extended (post-)Newtonian gravity algebra}

For $N=3$ we obtain the Newton--Hooke version of the algebra found in
\cite{Hansen:2018ofj} as the symmetry of post-Newtonian gravity. Moreover, when considering $N=4$, we find an extension of the 
post-Newtonian gravity algebra that generalises the algebra of \cite{Ozdemir:2019orp} to any dimension. 
Explicitly, for $N=3$ the generators of the expanded algebra are given by \eqref{expgens1}, \eqref{expgens2} and
\begin{equation}\label{expgens3}
\begin{aligned}
B_{a}=\lambda_{3}\otimes \tilde 
G_{a}\,,\\[5pt]
T_a=\lambda_{3}\otimes \tilde  P_{a}\,.
\end{aligned}
\end{equation}
 The zero of the semigroup $S_E^{(3)}$ is $\lambda_4$, and the reduction condition \eqref{redcon} together with the semigroup law and the (A)dS commutation relations \eqref{splitadsalg} leads to the following non-vanishing commutators for the expanded algebra:
\begin{multicols}{2} 
\begin{subequations}\label{ExtensionGalD2}
\setlength{\abovedisplayskip}{-15pt}
\allowdisplaybreaks
\begin{align}
[  G_{a},  H]&=P_{a}\, ,\\
[  G_{a},  M ]&=  T_{a}\, ,\\ 
[  B_{a},  H ]&=T_{a}\, ,\\ 
[  G_{a},  P_{b}]&=\delta_{ab}M\, , \\
[  G_{a},  G_{b}]&=S_{ab}\, ,\\
[  S_{ab},   G_{c}]&=2\delta_{c[b}   B_{a]}\, ,\\
[  S_{ab},  P_{c}]&=2\delta_{c[b}   T_{a]}\, ,\\
[  J_{ab},  J_{cd}]&=4\delta_{[a[c}  J_{d]b]}\, ,\\
[  J_{ab},  S_{cd}]&=4\delta_{[a[c}  S_{d]b]}\,,\\
[  P_{a},   H]&=\Lambda   G_{a}\, ,\\
[  P_{a},   M]&=\Lambda   B_{a}\, ,\\
[  T_{a},   H]&=\Lambda   B_{a}\, ,\\
[  P_{a},   P_{b}]&=-\Lambda   S_{ab}\, \\
[  J_{ab},  X_{c}]&=2\delta_{c[b}   X_{a]}\, ,\\
   X_{a}&= \left\{    G_{a},   P_{a},   B_{a},   T_{a} \right\} \,.\nonumber
\end{align}
\end{subequations}
\end{multicols}\noindent
In the case $\Lambda=0$ this algebra corresponds to the one found in \cite{Hansen:2018ofj} 
and further studied in~\cite{Bergshoeff:2019ctr,Gomis:2019fdh,Gomis:2019sqv}. 

One can extend this algebra by considering $N=4$ in the expansion prescription \eqref{redresp}. In this case, \eqref{expgens3} has to be supplemented with extra expanded generators given by
\begin{equation}\label{reds3}
Z_{ab}=\lambda_{4}\otimes \tilde  J_{ab} \,,\hspace{10mm}
Y=\lambda_{4}\otimes \tilde  H\,.
\end{equation}
The zero element in $S_E^{(4)}$ is given by $\lambda_5$, which means that the reduction condition \eqref{redcon} in this case sets
$
\lambda_5\otimes \tilde J_{AB}=\lambda_5\otimes \tilde P_{A} \equiv 0
$
and the non-vanishing commutation relations of the corresponding expanded algebra are given by \eqref{ExtensionGalD2} together with
\begin{multicols}{2} 
\begin{subequations}\label{ExtensionGalD3}
\setlength{\abovedisplayskip}{-15pt}
\allowdisplaybreaks
\begin{align}
[  G_{a},  T_{b}]&=\delta_{ab}Y\, , \\
[  B_{a},  P_{b}]&=\delta_{ab}Y\, , \\
[  G_{a},  B_{b}]&=Z_{ab}\, ,\\
[  J_{ab},  Z_{cd}]&=4\delta_{[a[c}  Z_{d]b]}\, ,\\
[  P_{a},   T_{b}]&=-\Lambda   Z_{ab}\,. \\
\end{align}
\end{subequations}
\end{multicols}\noindent
In the case $D=2+1$ and $\Lambda=0$ this algebra corresponds to the one studied in \cite{Ozdemir:2019orp} and defines a central extension of the post-Newtonian gravity symmetry \eqref{ExtensionGalD3}. For $D>2+1$, however, the generator $Z_{ab}$ is no longer central, exactly as it happens with $S_{ab}$ in the extension of the Newton--Hooke algebra.

 Notice that if some generators are central for some value $N$ of the semigroup, they are no longer central in the $N+1$ case.  More specifically, the expansions with odd $N$ do not possess central elements, while expansions with even values of $N$ always yield two central elements given by $\lambda_N\otimes \tilde J$ and tilde $\lambda_N\otimes \tilde H$.

Instead of continuing  with increasing values of $N$, in the next subsection we will 
construct an infinite-dimensional graded algebra
using an infinite-dimensional semigroup.

\subsubsection*{Infinite-dimensional Galilean algebra}

 The previous analysis suggests that we can construct an infinite-dimensional graded non-rela\-tivistic algebra using the expansion procedure
with an infinite-dimensional semigroup of the form \cite{Penafiel:2016ufo} 
 \begin{equation}\label{sinfty}
S^{(\infty)}=\left\{ \lambda_{0},\lambda_{1},\lambda_{2},\dots\right\} 
 \end{equation}
with multiplication law
 \begin{equation}
\lambda_{\alpha}\cdot\lambda_{\beta}=\lambda_{\alpha+\beta}\,.
 \end{equation}
(This semigroup can be realised as powers of a formal variable $\omega$ by letting $\lambda_\alpha=\omega^\alpha$. The finite semigroups $S^{(N)}_E$ then correspond to working to order $O(\omega^{N+1})$.)
As this semigroup does not have a zero element, the resonant subset decomposition that satisfies \eqref{resdecsemigroup} in this case is simply given by
\begin{equation}\label{infiniteS0S1}
\begin{aligned}
S_{0}^{(\infty)}&=\left\{ \lambda_{2m}\;|\;m=0,1,2\dots\right\}\,,\\
S_{1}^{(\infty)}&=\left\{ \lambda_{2m+1}\;|\;m=0,1,2\dots\right\}\,,
\end{aligned}
\end{equation}
and the corresponding resonant (non-reduced) expansion of (A)dS is
 \begin{equation}\label{redrespinf}
\Big\{S_{0}^{(\infty)}\otimes V_0\Big\} \oplus
\Big\{S_{1}^{(\infty)}\otimes V_1 \Big\}\,,
 \end{equation}
where $V_0$ and $V_1$ were defined in \eqref{splitAdS}. Denoting the generators of the expanded algebra \eqref{redrespinf} as
\begin{equation} \label{expgenkm}
\begin{aligned} %{ll}
  J^{(m)}_{ab}&=\lambda_{2m}\otimes \tilde  J_{ab}\,, &   B_{a}^{(m)}&=\lambda_{2m+1}\otimes \tilde  G_{a} \,,\\
  H^{(m)}&=\lambda_{2m}\otimes \tilde  H\,, &   P_{a}^{(m)}&=\lambda_{2m+1}\otimes \tilde  P_{a} \,,   
\end{aligned}
\end{equation}
leads to the infinite-dimensional graded algebra
\begin{multicols}{2}
\begin{subequations}\label{galkmalg}
\setlength{\abovedisplayskip}{-15pt}
\allowdisplaybreaks
\begin{align}
[  J^{(m)}_{ab},   P^{(n)}_{c}]&= 2\delta_{c[b}    P^{(m+n)}_{a]}\, ,\\
[  J^{(m)}_{ab},   J^{(n)}_{cd}]&=4\delta_{[a[c}    J^{(m+n)}_{d]b]}\, ,\\
[  B^{(m)}_{a},   P^{(n)}_{b}]&=  
\delta_{ab}    H^{(m+n+1)}\, , \\
[  B^{(m)}_{a},   H^{(n)}]&=   P^{(m+n)}_{a}\, ,\\
[  J^{(m)}_{ab},   B^{(n)}_{c}]&=2\delta_{c[b}    B^{(m+n)}_{a]}\, ,\\
[  B^{(m)}_{a},   B^{(n)}_{b}]&=    J^{(m+n+1)}_{ab}\,,\\
[  P^{(m)}_{a},   H^{(n)}]&=\Lambda   B^{(m+n)}_{a}\, ,\\
[  P^{(m)}_{a},   P^{(n)}_{b}]&= -\Lambda    J^{(m+n+1)}_{ab}\,.
\end{align}
\end{subequations}
\end{multicols}\noindent
We can get the finite-dimensional algebras presented in the previous subsections as quotients of \eqref{galkmalg} by suitable infinite ideals. 
For example, in the case of 
(\ref{ExtensionGalD2}) the ideal is generated by $J_{ab}^{(m)}, B_a^{(m)}, P_a^{(m)}, H^{(m)}$ for $m>2$.

Taking the limit $\Lambda \to 0$, we obtain as a contraction of \eqref{galkmalg}, the infinite-dimensional extension of the Galilei algebra
introduced in~\cite{Hansen:2019vqf} and further studied in~\cite{Gomis:2019fdh}.

\subsection{Extended Carroll (A)dS algebras}
\label{Carroll(A)dSexpansions}

The (A)dS Carroll algebra corresponds to the ultra-relativistic contraction of the (A)dS algebra \eqref{AdSalgCommutators}~\cite{Bacry:1968zf}. It can alternatively be obtained by means of an expansion procedure when considering the following subspace decomposition for (A)dS,
\begin{equation}
V_{0}=\{\tilde P_{a}, \tilde J_{ab}\}\,, \hspace{10mm}
V_{1}=\{\tilde H, \tilde G_{b}\}\,,\label{eq:splittingCarroll}
\end{equation}
which is $\mathbb{Z}_{2}$-graded. Note that here $P_a$ and $H$ have been interchanged with respect to the subspace decomposition used in the Newton--Hooke case \eqref{splitAdS}. This is a special
case of a general duality at the level of the translation generators between Galilean and Carrollian symmetries \cite{Barducci:2018wuj}. Using the decomposition \eqref{eq:splittingCarroll}, we can generalise the procedure outlined in the previous section to define Carrollian expansions of (A)dS, whose simplest case is the Carroll (A)dS algebra. Indeed, the non-vanishing commutation relations of the corresponding reduced resonant expanded algebra is given by
 \begin{equation}
\Big(s_{0}^{(N)}\times\{\tilde P_{a},\tilde J_{ab}\}
\Big)\oplus\Big(s_{1}^{(N)}\times\{\tilde H, \tilde G_{b}\} \Big)\,,
 \end{equation}
where $s_0^{(N)}$ and $s_1^{(N)}$ are given in \eqref{s0ands1}, and the case $N=1$ leads to the (A)dS Carroll algebra in $D$ dimensions
\begin{multicols}{2}
\begin{subequations}\label{carrollexps1}
\setlength{\abovedisplayskip}{-15pt}
\allowdisplaybreaks
\begin{align}
[  J_{ab},  P_{c}]&= 2\delta_{c[b}  P_{a]}\, ,\\
[  J_{ab},  J_{cd}]&=4\delta_{[a[c}  J_{d]b]}\, ,\\
[  G_{a},  P_{b}]&=\delta_{ab}H\, , \\
[  J_{ab},  G_{c}]&=2\delta_{c[b}  G_{a]}\, ,\\
[P_{a}, H]&=\Lambda G_{a}\, ,\\
[ P_{a}, P_{b}]&=-\Lambda  J_{ab}\, ,
\end{align}
\end{subequations}
\end{multicols}\noindent
where the expanded generators have been defined as
\begin{equation}
\begin{aligned}  
  P_a&=\lambda_{0}\otimes \tilde  P_{a}\,, &  H&=\lambda_{1}\otimes \tilde  H\,,
\\
   J_{ab}&=\lambda_{0}\otimes \tilde  J_{ab}\,,\quad &   G_{a}&=\lambda_{1}\otimes \tilde 
G_{a}\,.
\end{aligned}
\end{equation}
Naturally, when setting $\Lambda=0$, this allows one to obtain the Carroll algebra in $D$ dimensions as an expansion of the Poincar\'e algebra.  Subsequently, we can consider greater values of $N$ to obtain extended Carroll (A)dS algebras. In the case $N=2$ the expanded algebra has extra generators given by
 \begin{equation}
\begin{array}{ccc}
  T_a=\lambda_{2}\otimes \tilde  P_a\,,
\\
  S_{ab}=\lambda_{2}\otimes \tilde  J_{ab} \,.
\end{array}
 \end{equation}
and the commutation relations are the ones of \eqref{carrollexps1} together with 
\begin{multicols}{2}
\begin{subequations}\label{carrollexps2}
\setlength{\abovedisplayskip}{-15pt}
\allowdisplaybreaks
\begin{align}
[  G_{a},  H]&=T_{a}\, ,\\
[  G_{a},  G_{b}]&=S_{ab}\, ,\\
[  J_{ab},  T_{c}]&= 2\delta_{c[b}  T_{a]}\, ,\\
[  S_{ab},  P_{c}]&= 2\delta_{c[b}  T_{a]}\, ,\\
[  J_{ab},  S_{cd}]&=4\delta_{[a[c}  S_{d]b]}\,\\
[ P_{a}, T_{b}]&=-\Lambda  S_{ab}\, .
\end{align}
\end{subequations}
\end{multicols}\noindent
In the case $N=3$, we get two extra generators,
\begin{equation}
\begin{aligned}
&   M=\lambda_3 \otimes \tilde  H\,,
\\
&   B_a=\lambda_3 \otimes \tilde  G_{a}\,.
\end{aligned}
\end{equation}
The commutation relations of this expanded algebra are \eqref{carrollexps1}, \eqref{carrollexps2} plus
\begin{multicols}{2}
\begin{subequations}\label{carrollexps3}
\setlength{\abovedisplayskip}{-15pt}
\allowdisplaybreaks
\begin{align}
[  J_{ab},  B_{c}]&= 2\delta_{c[b}  B_{a]}\, ,\\
[  G_{a},  T_{b}]&=\delta_{ab}M\, , \\
[  B_{a},  P_{b}]&=\delta_{ab}M\, , \\
[  S_{ab},  G_{c}]&=2\delta_{c}  B_{a]}\, ,\\
[T_{a}, H]&=\Lambda B_{a}\, ,\\
[P_{a}, M]&=\Lambda B_{a}\, .
\end{align}
\end{subequations}
\end{multicols}\noindent
For $\Lambda=0$, this algebra defines a Carrollian counterpart of the (post-)Newtonian symmetry introduced in \cite{Hansen:2019vqf} in the context of general relativity.
In the same way, one can check that the $N=4$ case defines a higher-dimensional Carrollian (A)dS analogue of the extended \mbox{(post-)}Newtonian symmetry given in \cite{Ozdemir:2019orp}.

\subsubsection*{Infinite-dimensional Carroll algebra}
Similarly to the Newton--Hooke construction, the ultra-relativistic expansions of (A)dS form a family of algebras, which can be described in a unified fashion by considering the infinite semigroup \eqref{sinfty} and the non-reduced resonant expansion 
 \begin{equation}
\label{eq:Carrollinfty}
\Big(S_{0}^{(\infty)}\times\left\{ \tilde{J}_{ab},\tilde{P}_a \right\} \Big)\oplus\Big(S_{1}^{(\infty)}\times\left\{ \tilde{G}_{a},\tilde{H}\right\} \Big)\,,
 \end{equation}
where $S_0^{(\infty)}$ and $S_1^{(\infty)}$ are given in \eqref{infiniteS0S1}. By defining an infinite set of expanded generators by
\begin{equation}\label{expgencarrollkm}
\begin{aligned}
  J^{(m)}_{ab}&=\lambda_{2m}\otimes \tilde  J_{ab}\,, &   B _{a}^{(m)}&=\lambda_{2m+1}\otimes \tilde  G_{a} \,,\\
  P_{a}^{(m)}&=\lambda_{2m}\otimes \tilde  P_{a} \,, &   H^{(m)}&=\lambda_{2m+1}\otimes \tilde  H\,,
\end{aligned}
\end{equation}
we obtain an infinite-dimensional Carrollian expansion of the (A)dS algebra
\begin{multicols}{2}
\begin{subequations}\label{carrollkmalg}
\setlength{\abovedisplayskip}{-15pt}
\allowdisplaybreaks
\begin{align}
[  J^{(m)}_{ab},   P^{(n)}_{c}]&= 2\delta_{c[b}    P^{(m+n)}_{a]}\, ,\\
[  J^{(m)}_{ab},   J^{(n)}_{cd}]&=4\delta_{[a[c}    J^{(m+n)}_{d]b]}\, ,\\
[  B^{(m)}_{a},   P^{(n)}_{b}]&=\delta_{ab}    H^{(m+n)}\, , 
\\
[  B^{(m)}_{a},   H^{(n)}]&=P^{(m+n+1)}_{a}\, ,
\\
[  J^{(m)}_{ab},   B^{(n)}_{c}]&=2\delta_{c[b}    B^{(m+n)}_{a]}\, ,\\
[  B^{(m)}_{a},   B^{(n)}_{b}]&=    J^{(m+n+1)}_{ab}\,.
\\
[ P_{a}^{(m)}, H^{(n)}]&=\Lambda  B_{a}^{(m+n)}\, ,\\
[ P_{a}^{(m)}, P_{b}^{(n)}]&=-\Lambda  J_{ab}^{(m+n)}\, .
\end{align}
\end{subequations}
\end{multicols}\noindent
The different finite expansions previously constructed using the semigroup $S_E^{(N)}$ can be obtained from the infinite case by considering suitable quotients. Redefining the generators according to
\begin{align}
J_{ab}^{(m)} &\to J_{ab}^{(m)}\,,\nn\\
P_{a}^{(m)} &\to P_{a}^{(m-1)}\quad (m\geq1)\,,\nn\\
B_{a}^{(m)} &\to B_{a}^{(m)}\,,\nn\\
H^{(m)} &\to - H^{(m)} 
\end{align}
and, 
taking the limit $\Lambda\to 0$, we obtain an extension of the
infinite-dimensional algebra in~\cite{Hansen:2019vqf} as a contraction of the infinite-dimensional extended Carrollian algebra \eqref{carrollkmalg}. A related contraction was obtained from the infinite-dimensional extended Newton--Hooke algebra \eqref{galkmalg} above.
In the contraction of the Carroll algebra \eqref{carrollkmalg},
it is extended in a semidirect sum by the additional generator $P_{a}^{(0)}$ (which should then rather be called $P_{a}^{(-1)}$ in the notation of~\cite{Hansen:2019vqf}).
Conversely, the algebra in \cite{Hansen:2019vqf} can be seen as an extension of the contracted infinite-dimensional extended Carroll algebra
if we again redefine $H^{(m)} \to H^{(m+1)}$ and then add a generator $H^{(0)}$.
\section{Newton--Hooke and Carrollian affine algebra}
\label{sectionNHfreealgebra}

In this section, we will show that the infinite-dimensional Lie algebras (\ref{galkmalg}) and~\eqref{carrollkmalg} also can be obtained from 
the extension of $\mathfrak{so}(D-1)$ to
an (untwisted or twisted) affine Kac--Moody algebra. We shall only require parabolic subalgebras of these Kac--Moody algebras and this construction can be linked to free Lie algebras in terms of quotients by Serre relations. The algebras for finite $N$ discussed in the previous section then correspond to further quotients, similar to the constructions in~\cite{Gomis:2017cmt,Gomis:2019fdh}.

\subsection{Newton--Hooke affine algebras}

\begin{figure}[t]
\centering
\begin{picture}(227,87.2)
\thicklines
\put(10,20){\circle{7}}
\put(13.5,20){\line(1,0){29.8}}
\put(47,20){\circle{7}}
\put(47,20){\circle{7}}
\put(184,20){\circle{7}}
\put(47,23.5){\line(0,1){30.2}}
\put(184,23.5){\line(0,1){30.2}}
\put(184,57.2){\circle{7}}
\put(47,57.2){\circle{7}}
\put(187.7,20){\line(1,0){30.2}}
\put(221.7,20){\circle{7}}
\put(50.5,20){\line(1,0){30.2}}
\put(180.4,20){\line(-1,0){30.2}}
\put(84.1,20){\circle{7}}
\put(146.3,20){\circle{7}}
\multiput(87.5,20)(22,0){3}{\line(1,0){11}}
\put(8,5){\scriptsize$0$}
\put(45,5){\scriptsize$1$}
\put(213,5){\scriptsize$r-1$}
\put(195,55){\scriptsize$r$}
\put(58,55){\scriptsize$-1$}
\end{picture}
\begin{picture}(227,87.2)
\thicklines
\put(58,55){\scriptsize$-1$}
\put(47,57.2){\circle{7}}
\put(47,23.5){\line(0,1){30.2}}
\put(10,20){\circle{7}}
\put(13.5,20){\line(1,0){29.8}}
\put(47,20){\circle{7}}
\put(218.2,20){\line(-1,1){5}}
\put(218.2,20){\line(-1,-1){5}}
\put(184,20){\circle{7}}
\put(187.7,21){\line(1,0){29.2}}
\put(187.7,19){\line(1,0){29.2}}
\put(221.7,20){\circle{7}}
\put(50.5,20){\line(1,0){30.2}}
\put(180.4,20){\line(-1,0){30.2}}
\put(84.1,20){\circle{7}}
\put(146.3,20){\circle{7}}
\multiput(87.5,20)(22,0){3}{\line(1,0){11}}
\put(7.5,5){\scriptsize$0$}
\put(44,5){\scriptsize$1$}
\put(220,5){\scriptsize$r$}
\end{picture}
\caption{
\it Dynkin diagrams of $B_{r+1}^{(1)}$ (lower) and $D_{r+1}^{(1)}$ (upper). These are relevant for the infinite Newton--Hooke algebras.}\label{fig:Br2}
\end{figure}
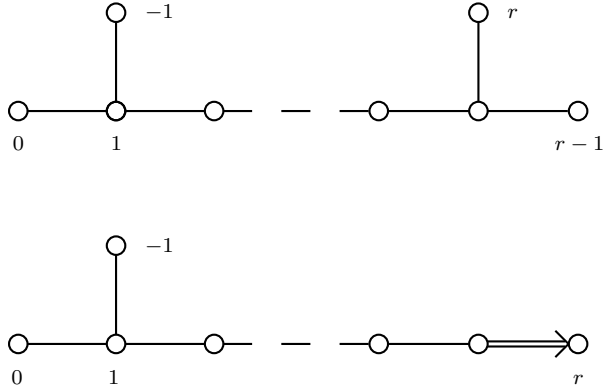

Consider first the complex Lie algebra
$\mathfrak{g}_r=D_r$
if $d=D-1$ is even ($d=2r$), or $\mathfrak{g}_r=B_r$ if $d=D-1$ is odd ($d=2r+1$). Thus 
$\fg_r$ is the complexification of $\so(d)$.
We can extend $\fg_r$ to $\fg_{r+1}$ (either $D_{r+1}$ or $B_{r+1}$) by adding a node labelled $0$ and then further to an affine Kac--Moody algebra $\fg_{r+1}^{(1)}$ by adding
a node labelled $-1$ to the Dynkin diagram of $\fg_r$. The resulting diagram, together with our labelling of the nodes, is shown in Figure \ref{fig:Br2}.
The corresponding Cartan matrix is
\setlength\arraycolsep{10pt}
\renewcommand\arraystretch{1}
\begin{align}
A_{ij}=
\begin{pmatrix}
2 & 0 & -1 &  \cdots & 0 & 0& 0\\ 
0 & 2 & -1 &  \cdots & 0 & 0& 0\\ 
-1 & -1 & 2&  \cdots & 0 & 0& 0\\ 
\vdots &  \vdots &  \vdots &  \ddots & \vdots & \vdots& \vdots  \\
0 & 0 & 0 & \cdots & 2 & -1 & -1\\
0 & 0 & 0 & \cdots & -1 & 2 & 0\\
0 & 0 & 0 & \cdots & -1 & 0 & 2
\end{pmatrix}
\end{align} 
for $D_{r+1}^{(1)}$
and
\setlength\arraycolsep{10pt}
\renewcommand\arraystretch{1}
\begin{align}
A_{ij}=
\begin{pmatrix}
2 & -2 & 0 &  \cdots & 0 & 0 & 0\\ 
-1 & 2 & -1 &  \cdots & 0 & 0 & 0\\ 
0 & -1 & 2&  \cdots & 0 & 0 & 0\\ 
\vdots &  \vdots &  \vdots &  \ddots & \vdots & \vdots & \vdots  \\
0 & 0 & 0 & \cdots &2& -1 & 0\\
0 & 0 & 0 & \cdots &-1& 2 & -1\\
0 & 0 & 0 & \cdots &0& -2 & 2
\end{pmatrix}
\end{align} 
for $B_{r+1}^{(1)}$ (where rows and columns are counted $-1,0,1,\ldots,r$ from left to right and from top to bottom).

When we add the two nodes $0$ and $-1$ we also add six generators $e_i,f_i,h_i$ for $i=0,-1$ to the generators of $\fg_r$. 
This gives a central extension of the loop algebra of $\fg_{r+1}$ where the central element is given by
\begin{align}
c=h_{-1}+h_0+2h_1+2h_2+\cdots+2h_{r-2}+h_{r-1}+h_r\,.
\end{align}
The affine algebra $\fg_{r+1}^{(1)}$ is then obtained from it by adding also a derivation generator $d$ satisfying $[d,e_{-1}]=e_{-1}$ and $[d,f_{-1}]=-f_{-1}$
and commuting with all the other generators.

We consider the subalgebra of
$\fg_{r+1}^{(1)}$ generated by $e_i,h_i$ for $i=0,-1$ together with the generators of $\fg_r$. This Lie algebra has an $(\mathbb{N}\times \mathbb{N})$-grading associated to these two nodes, which can be trivially extended to a $(\mathbb{Z}\times \mathbb{Z})$-grading.
It can thus be decomposed into a direct sum of subspaces, each labelled by a pair $(\ell_0,\ell_{-1})$ of non-negative integers. The subspace labelled by $(\ell_0,\ell_{-1})$
is spanned by elements formed as multibrackets of the generators, where $e_0$ and $e_{-1}$ appear $\ell_0$ and $\ell_{-1}$ times, respectively.
The subalgebra at $(\ell_0,\ell_{-1})=(0,0)$ contains $\fg_r$ but also the two additional Cartan generators $h_{0}$ and $h_{-1}$. By taking the linear combinations
\begin{equation}
\begin{aligned}
&h=h_{-1}+h_1+h_2+\cdots+h_{r-2}+\tfrac12 h_{r-1}+\tfrac12 h_r\,,\\
&h'=h_{0}+h_1+h_2+\cdots+h_{r-2}+\tfrac12 h_{r-1}+\tfrac12 h_r\,,
\end{aligned}
\end{equation}
(such that $c=h+h'$) we get elements that commute with $\fg_r$. Furthermore, they satisfy
\begin{equation} \label{eigenvalues}
\begin{aligned}
&[h,e_{-1}]=e_{-1}\,,&& [h',e_{-1}]=-e_{-1}\,,\\
&[h,e_{0}]=-e_{0}\,,&& [h',e_{0}]=e_{0}\,.
\end{aligned}
\end{equation}
The $(\mathbb{Z}\times \mathbb{Z})$-grading gives rise to a $\mathbb{Z}$-grading, where the single level $\ell$ is the sum of $\ell_0$ and $\ell_{-1}$.
Since $\fg_r$ is a subalgebra at level zero, we get representations of it at each level.

At level $\ell=0$ we have the generators $J_{ab}$ of (the complexification of) $\mathfrak{so}(d)$ and the two $\mathfrak{so}(d)$ scalars $h$ and $h'$.
At level $\ell=1$ we have two lowest weight representations
with lowest weight vectors
$e_{-1}$ and $e_0$, respectively. 
The Dynkin labels of both corresponding highest weight representations are $[1,0,0,\ldots,0]$ since
\begin{equation}
\begin{aligned}
&[h_1,e_{-1}]=-1\,, \qquad&& [h_2,e_{-1}]=[h_2,e_{-1}]=\cdots= [h_r,e_{-1}]=0\,,\\
&[h_1,e_{0}]=-1\,, \qquad && [h_2,e_{0}]=[h_2,e_{0}]=\cdots= [h_r,e_{0}]=0\,.
\end{aligned}
\end{equation}
Thus they are vector representations, and we denote the corresponding generators by $X_a$ (with lowest weight vector $e_{-1}$) and $Y_a$
(with lowest weight vector $e_{0}$), where $a=1,2,\ldots,d$ as before.
It follows from (\ref{eigenvalues}) that
\begin{equation}
\begin{aligned}
&[h,X_a]=X_a\,,\qquad&& [h',X_a]=-X_a\,,\\
&[h,Y_a]=-Y_a\,,\qquad&& [h',Y_a]=Y_a\,.
\end{aligned}
\end{equation}
(The parabolic subalgebra of the `horizontal' algebra $\mathfrak{g}_{r+1}$ is generated by $J_{ab}$, $Y_a$ and $h_0$.)

In the {\it free Lie algebra} generated by all $X_a$ and $Y_a$ at level $\ell=1$, the subspace at level $\ell=2$ decomposes into a direct sum of $\so(d)$
modules with Dynkin labels
\begin{equation}
\begin{aligned}
2[1,0,0,\ldots,0] \wedge 2[1,0,0,\ldots,0] &= 2([1,0,0,\ldots,0] \wedge [1,0,0,\ldots,0])\\
&\quad\, \oplus [1,0,0,\ldots,0]\otimes [1,0,0,\ldots,0]\\
&=3 [0,1,0,\ldots,0]\oplus[2,0,0,\ldots,0]\oplus[0,0,0,\ldots,0]\, ,\label{antisymprod}
\end{aligned}
\end{equation}
(where $\wedge$ denotes the antisymmetric tensor product, coming from the antisymmetry of the Lie bracket). The free Lie algebra construction continues to all positive levels~\cite{Gomis:2017cmt,Gomis:2019fdh} but in order to reproduce the algebra~\eqref{galkmalg} we need to take a quotient.
This quotient leads to the subalgebra of $\fg_{r+1}^{(1)}$ at positive levels and the ideal that one quotients out is generated by
the {\it Serre relations}
\begin{align}
[e_{-1},[e_{-1},e_1]]=[e_{0},[e_{0},e_1]]=[e_{-1},e_0]=0\,,
\end{align}
corresponding to the representation
\begin{align}
2[0,1,0,\ldots,0]\oplus [2,0,0,\ldots,0]\,.
\end{align}
Thus, this representation has to be removed from the antisymmetric tensor product (\ref{antisymprod}), leaving only
the direct sum
\begin{align}
[0,1,0,\ldots,0]\oplus[0,0,0,\ldots,0]
\end{align}
at level $\ell=2$. We denote the corresponding 2-form and scalar generators by $J_{ab}{}^{2}$ and $h^{2}$, respectively, where the superscript indicates that they appear at level 
$\ell=2$.
The commutation relations among the generators at level $\ell=1$ giving rise to these generators at level $\ell=2$ are
\begin{align}
[X_a,Y_b]&=J_{ab}{}^{2} + \delta_{ab}h^2\,,& [X_a,X_b]&=[Y_a,Y_b]=0\,.
\end{align}

The pattern with two vectors at odd levels and a 2-form and a scalar at even levels continues, as shown in Table \ref{affinetable}. The set of generators at non-negative levels are\footnote{In this notation, we have singled out $h'$ arbitrarily. As the diagrams are symmetric under the exchange of nodes $0$ and $-1$, we could have also exchanged the roles of $h$ and $h'$.}
\begin{align}
\{h',h^{2k},J_{ab}{}^{2k},X_a{}^{2k+1},Y_a{}^{2k+1}\}
\end{align}
for $k\geq 0$ (where again the superscript is the level $\ell$), and the 
non-vanishing commutation relations (except for those involving $h'$, which we have omitted since they turn out 
to be irrelevant) are
\begin{align}
[{J}_{ab}{}^{2k},{J}_{cd}{}^{2k'}]&=4\delta_{[c[b}{J}_{a]d]}{}^{2(k+k')}\,,\nn\\
[{J}_{ab}{}^{2k},{X}_c{}^{2k'+1}]&=2\delta_{c[b}{X}_{a]}{}^{2(k+k')+1}\,,\nn\\
[h{}^{2k},{X}_a{}^{(2k'+1)}]&={X}_{a}{}^{2(k+k')+1}\,,\nn\\
[{J}_{ab}{}^{2k},{Y}_c{}^{2k'+1}]&=2\delta_{c[b}{Y}_{a]}{}^{2(k+k')+1}\,,\\
[h{}^{2k},{Y}_a{}^{2k'+1}]&=-{Y}_{a}{}^{2(k+k')+1}\,,\nn\\
[{X}_a{}^{2k+1},{Y}_b{}^{2k'+1}]&=J_{ab}{}^{2(k+k'+1)} + \delta_{ab}h{}^{2(k+k'+1)}\,,\nn
\end{align}
where ${J}_{ab}{}^{0}={J}_{ab}$, $h{}^{0}=h$, $X_a{}^{1}=X_a$ and $Y_a{}^{1}=Y_a$.
If we then set
\begin{equation}
\begin{aligned}
&J_{ab} =J_{ab}{}^{0}\,,\qquad
&& S_{ab} =J_{ab}{}^{2}\,,\\
&H =\sqrt{\Lambda}\,h{}^{0}\,,\qquad
&& M =\sqrt{\Lambda}\,h{}^{2}\,,\\
&G_a =\frac{1}{\sqrt{2}}(Y_a{}^{1}+X_a{}^{1})\,,\qquad
&& B_a=\frac{1}{\sqrt{2}}(Y_a{}^{3}+X_a{}^{3})\,,\\
&P_a =\frac{\sqrt{\Lambda}}{\sqrt{2}}(Y_a{}^{1}-X_a{}^{1})\,,\qquad
&& T_a =\frac{\sqrt{\Lambda}}{\sqrt{2}}(Y_a{}^{3}-X_a{}^{3})\,,
\end{aligned}
\end{equation}
for $\Lambda > 0$, then we recover the commutation relations (\ref{ExtensionGalD2}). More generally, if we set
\begin{equation}
\begin{aligned}
&J_{ab}{}^{(m)} =J_{ab}{}^{2m}\,,\\
&H{}^{(m)} =\sqrt{\Lambda}\,h{}^{2m}\,,\\
&B_a{}^{(m)} =\frac1{\sqrt{2}}(Y_a{}^{2m+1}+X_a{}^{2m+1})\,,\\
&P_a{}^{(m)} =\frac{\sqrt{\Lambda}}{\sqrt{2}}(Y_a{}^{2m+1}-X_a{}^{2m+1})\,
\end{aligned}
\end{equation}
for $\Lambda > 0$, then we recover the commutation relations (\ref{galkmalg}). These formulas are still valid for $\Lambda<0$ if we interpret
$\sqrt{\Lambda}$ as $\pm i\sqrt{|\Lambda|}$ and we will then obtain a different real form of the complex Lie algebra $\fg_{r+1}^{(1)}$.
\begin{table}[t]
\setlength{\arraycolsep}{10pt}
\renewcommand{\arraystretch}{1.5}
\begin{align*}
\begin{array}{c|c|c|c|c|c|c}
&\ell_{-1}=0 & \ell_{-1}=1 &\ell_{-1}=2 & \ell_{-1}=3 &\ell_{-1}=4 & \cdots\\
\hline
\ell_{0}=0 & \!\!\!\!J_{ab}{}^0\,,\, h^0\,,\,h'\!\!\!\! & X_a{}^1 &&&&\\
\hline
\ell_{0}=1 & Y_{a}{}^1 & \,\,J_{ab}{}^2\,,\, h^2\,\, &X_a{}^3&&&\\
\hline
\ell_{0}=2 &  & Y_{a}{}^3&\,\,J_{ab}{}^4\,,\, h^4\,\,&X_a{}^5&&\\
\hline
\ell_{0}=3 & & &Y_a{}^5&\,\,J_{ab}{}^6\,,\, h^6\,\,&X_a{}^7&\\
\hline
\ell_{0}=4 & & &&Y_{a}{}^7 &\,\,J_{ab}{}^8\,,\, h^8\,\,&\ddots\\
\hline
\vdots & &  &&&\ddots& \ddots
\end{array}
\end{align*}\caption{\it The non-negative levels of the affine Kac--Moody algebra $\fg_{r+1}^{(1)}$.}\label{affinetable}
\end{table}

\subsection{(A)dS Carrollian affine algebras}

The procedure for obtaining~\eqref{carrollkmalg} from an affine algebra is very similar to the previous discussion so we shall be rather brief.

We first note that due to the choice of subspaces in~\eqref{eq:splittingCarroll}, the infinitely expanded algebra~\eqref{carrollkmalg} consists of infinite repetitions of copies of $\{\tilde{J}_{ab}, \tilde{P}_a\}$ as even level spaces and $\{\tilde{H}, \tilde{G}_a\}$ as odd level spaces, see~\eqref{eq:Carrollinfty}. Since in the Carrollian limit the commutation relations~\eqref{carrollexps1} are perfectly compatible with identifying the complexification of $V_0=\{\tilde{J}_{ab}, \tilde{P}_a\}$ as the algebra $\mathfrak{so}(D)$ and
the complexification of $V_1=\{\tilde{H}, \tilde{G}_a\}$ as its $D$-dimensional vector representation, we have to look for a Kac--Moody algebra where these two spaces repeat infinitely. 
Thus, there must be Serre relations corresponding to an ideal
such that we obtain the positive levels as a quotient of the free Lie algebra generated by the space $V_1$.

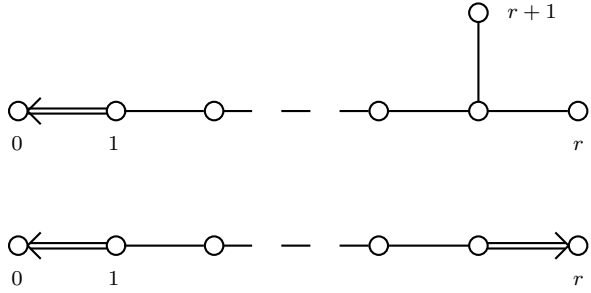
\begin{figure}[t!]
\centering
\begin{picture}(227,87.2)
\thicklines
\put(10,20){\circle{7}}
\put(13.5,21){\line(1,0){30.2}}
\put(13.5,19){\line(1,0){30.2}}
\put(13.5,20){\line(1,1){5}}
\put(13.5,20){\line(1,-1){5}}
\put(47,20){\circle{7}}
\put(184,20){\circle{7}}
\put(184,23.5){\line(0,1){30.2}}
\put(184,57.2){\circle{7}}
\put(187.7,20){\line(1,0){30.2}}
\put(221.7,20){\circle{7}}
\put(50,20){\line(1,0){30.2}}
\put(180.4,20){\line(-1,0){30.2}}
\put(84.1,20){\circle{7}}
\put(146.3,20){\circle{7}}
\multiput(87.5,20)(22,0){3}{\line(1,0){11}}
\put(7.5,5){\scriptsize$0$}
\put(44,5){\scriptsize$1$}
\put(220,5){\scriptsize$r$}
\put(195,55){\scriptsize$r+1$}
\end{picture}
\begin{picture}(227,50)
\thicklines
\put(10,20){\circle{7}}
\put(14.5,21){\line(1,0){29.2}}
\put(14.5,19){\line(1,0){29.2}}
\put(47,20){\circle{7}}
\put(13.5,20){\line(1,1){5}}
\put(13.5,20){\line(1,-1){5}}
\put(218.2,20){\line(-1,1){5}}
\put(218.2,20){\line(-1,-1){5}}
\put(184,20){\circle{7}}
\put(187.7,21){\line(1,0){29.2}}
\put(187.7,19){\line(1,0){29.2}}
\put(221.7,20){\circle{7}}
\put(50,20){\line(1,0){30.2}}
\put(180.4,20){\line(-1,0){30.2}}
\put(84.1,20){\circle{7}}
\put(146.3,20){\circle{7}}
\multiput(87.5,20)(22,0){3}{\line(1,0){11}}
\put(7.5,5){\scriptsize$0$}
\put(44,5){\scriptsize$1$}
\put(220,5){\scriptsize$r$}
\end{picture}
\caption{
\it Dynkin diagrams of $B_{r+1}^{(1)}$ (upper) and $D_{r+1}^{(2)}$ (lower). These are relevant to the (A)dS Carroll algebras.}\label{fig:carrollkm}
\end{figure}

The problem turns out to be identical to one already solved in~\cite{Gomis:2019fdh}, where the corresponding algebras were identified as the untwisted affine algebra of type 
$B_{r+1}^{(1)}$ (for $D$ even and $r+1=D/2$) and as the twisted affine algebra $D_{r+1}^{(2)}$ (for $D$ odd and $r=(D-1)/2$). The corresponding Dynkin diagrams are reproduced in Figure~\ref{fig:carrollkm} for convenience.  By 
redefining the generators in \eqref{carrollkmalg} according to
\begin{align}
J_{ab}^{(m)} &\to J_{ab}^{(m)}\,,\nn\\
P_{a}^{(m)} &\to \sqrt{-\Lambda}\, P_{a}^{(m-1)}\quad (m\geq1)\,,\nn\\
B_{a}^{(m)} &\to B_{a}^{(m)}\,,\nn\\
H^{(m)} &\to -\sqrt{-\Lambda}\, H^{(m)} ,
\end{align}
in the AdS case ($\Lambda <0$)
we recover the commutation relations in (3.11) of~\cite{Gomis:2019fdh}. 
The dS case ($\Lambda >0$) corresponds to taking a different real form of the complex Kac--Moody algebra, interpreting
$\sqrt{-\Lambda}$ as $\pm i \sqrt{\Lambda}$.

\section{Chern--Simons non-relativistic gravities}

In this section we want to find realisations of the previous non-relativistic symmetries. We will be interested in the construction of Chern--Simons gravities in $2+1$ dimensions, defined by the action\footnote{All our integrals are over three-dimensional space-time.}
\begin{equation}\label{csactiongalinf}
S_{\text{CS}}[A]=\int\left\langle A\wedge dA+\frac{2}{3}A\wedge A\wedge A\right\rangle \,,
\end{equation}
where the gauge algebra is given by \eqref{galkmalg} and~\eqref{carrollkmalg}. In order to carry out this construction we shall construct invariant tensors of these algebras in the following way: in $2+1$ dimensions, an invariant tensor for the expanded algebras of interest can be obtained following \cite{izaurieta:2006zz}, by first defining the structure constants $K_{\;\;\alpha\beta}^{\gamma}$ of the semigroup \eqref{sinfty} as
\begin{equation}
\lambda_\alpha \lambda_\beta = K_{\;\;\alpha\beta}^{\gamma}\,\lambda_\gamma\,.
\end{equation}
Given the invariant tensor on $\mathfrak{so}(2,2)$,
\begin{equation}
\label{eq:so22inv1}
\left\langle \tilde J_{AB} \tilde P_{C} \right\rangle = \epsilon_{ABC}\quad (\epsilon_{012}=-1)\,,
\end{equation}
the invariant tensor on the expanded algebra can be defined as
\begin{equation}\label{expandedinvtensor}
\left\langle \left(\lambda_\alpha\otimes \tilde  J_{AB}\right) \left( \lambda_\beta \otimes \tilde  P_{C}\right) \right\rangle = \alpha_{(\gamma)} K_{\;\;\alpha\beta}^{\gamma} \,\epsilon_{ABC}\,,
\end{equation}
where we have introduced an infinite set of arbitrary constants
$\alpha_{(\gamma)}$.

\subsection{Extended Newton--Hooke gravities}
Now we will consider the infinite-dimensional extended Newton--Hooke algebra \eqref{galkmalg}, where the definition \eqref{expgenkm} of the expanded generators leads to 
\begin{align}
\left\langle B_{a}^{(m)}P_{b}^{(n)}\right\rangle &=-\alpha_{(2m+2n+2)}\epsilon_{ab}\,, &  
\left\langle J_{ab}^{(m)}H^{(n)}\right\rangle &=-\alpha_{(2m+2n)}\epsilon_{ab}\,.  \label{invtensor0}
\end{align}
Therefore, as the constants $\alpha_{(\gamma)}$ in \eqref{invtensor0} are non-vanishing only for even values of $\gamma$, it is convenient to relabel them in terms of a set of constants $\mu_{(m)}=\alpha_{(2m)}$. Finally, using \eqref{expgenkm} and dualising $J_{ab}$ and $B_a$ in the form
\begin{align}
J^{(m)}&\equiv\frac{1}{2}\epsilon^{ab}J^{(m)}_{ab}\,,& G^{(m)}_a&\equiv\epsilon_{a}^{\;\;b}B^{(m)}_b\,,  \label{redefJP}  
\end{align}
($a,b=1,2$) yields 
\begin{align}\label{invtensor}
\left\langle G_{a}^{(m)}P_{b}^{(n)}\right\rangle &=\mu_{(m+n+1)}\delta_{ab}\,,&\left\langle J^{(m)}H^{(n)}\right\rangle &=-\mu_{(m+n)}\,.
\end{align}
In terms of the dual generators \eqref{redefJP}, the infinite-dimensional graded algebra \eqref{galkmalg} takes the form
\begin{multicols}{2}
\begin{subequations}\label{galkmalg2+1}
\setlength{\abovedisplayskip}{-15pt}
\allowdisplaybreaks
\begin{align}
[J^{(m)},P_{a}^{(n)}] &=-\epsilon_{a}^{\;\;b}P_{b}^{(m+n)}\,,\\
[G_{a}^{(m)},G_{b}^{(n)}] &=\epsilon_{ab}J^{(m+n+1)}\,,\\
[J^{(m)},G_{a}^{(n)}] &=-\epsilon_{a}^{\;\;b}G_{b}^{(m+n)}\,,\\
[G_{a}^{(m)},P_{b}^{(n)}] &=\epsilon_{ab}H^{(m+n+1)}\,,\\
[H^{(m)},G_{a}^{(n)}] &=-\epsilon_{a}^{\;\;b}P_{b}^{(m+n)}\,,\\
[H^{(m)},P_{a}^{(n)}] &=\Lambda \epsilon_{a}^{\;\;b}G_{b}^{(m+n)}\,,\\
[P_{a}^{(m)},P_{b}^{(n)}] &=-\Lambda\epsilon_{ab}J^{(m+n+1)}\,.
\end{align}
\end{subequations}
\end{multicols}\noindent

\subsubsection*{Action}
In order to construct a Chern--Simons action, we define a connection one-form taking values on \eqref{galkmalg2+1}
\begin{equation}\label{csconnection}
A=\sum_{m=0}^{\infty}\left(e_{(m)}^{a}P_{a}^{(m)}+\omega_{(m)}^{a}G_{a}^{(m)}+\tau_{(m)} H^{(m)} +\omega_{(m)}J^{(m)}\right)\,.
\end{equation}
The curvature 2-form $F=dA+A\wedge A$ then reads
\begin{equation} \label{curvatureform}
F=\sum_{m=0}^{\infty}\Bigg(F^{a}\left[P^{(m)}\right]P_{a}^{(m)}+F^{a}\left[G^{(m)}\right]G_{a}^{(m)}
+F\left[H^{(m)}\right]H^{(m)}+F\left[J^{(m)}\right]J^{(m)}\Bigg)\,.
\end{equation}
where the curvature components can be worked out using the algebra~\eqref{galkmalg2+1} and read explicitly 
\begin{equation}\label{curvatureform2}
\begin{aligned}
F^{a}\left[P^{(m)}\right] & =de_{(m)}^{a}-\sum_{n,p=0}^{\infty}\delta_{m}^{n+p}\,\epsilon_{\;b}^{a}\left(\omega_{(n)}^{b}\tau_{(p)}+ e_{(n)}^{b}\omega_{(p)} \right)\\
F^{a}\left[G^{(m)}\right] & =d\omega_{(m)}^{a}-\sum_{n,p=0}^{\infty}\delta_{m}^{n+p}\,\epsilon_{\;b}^{a}\; \left(\omega_{(n)}^{b}\omega_{(p)}-\Lambda e_{(n)}^{b}\tau_{(p)}\right) \,,\\
F\left[H^{(m)}\right] & =d\tau_{(m)}+\sum_{n,p=0}^{\infty}\delta_{m}^{n+p+1}\,\epsilon_{ab}\;e_{(n)}^{a}\omega_{(p)}^{b}\,,\\
F\left[J^{(m)}\right] & =d\omega_{(m)}+\frac{1}{2}\sum_{n,p=0}^{\infty}\delta_{m}^{n+p+1}\,\epsilon_{ab}\;\left(\omega_{(n)}^{a}\omega_{(p)}^{b}-\Lambda\,e_{(n)}^{a}e_{(p)}^{b}\right)\,.\\
\end{aligned}
\end{equation}
Here and from now on the wedge product between forms are not written explicitly. Using the gauge connection \eqref{csconnection} and the invariant bilinear form \eqref{invtensor}, the action \eqref{csactiongalinf} takes the form
\begin{equation}\label{csactiongalinf2}
\begin{aligned}
S_{\text{CS}} =&\sum_{m,n=0}^{\infty}\mu_{(m+n+1)}\int \Bigg( e_{(m)}^{a}F_a\left[G^{(n)}\right]+\omega_{(m)}^{a}F_a\left[P^{(n)}\right]\Bigg)\\
-&\sum_{m,n=0}^{\infty}\mu_{(m+n)}\int \Bigg(\tau_{(m)}F\left[J^{(n)}\right]+\omega_{(m)}F\left[H^{(n)}\right]\Bigg)\\
+&\sum_{m,n,p=0}^{\infty}\mu_{(m+n+p+1)}\,\int \epsilon_{ab}\left(e_{(m)}^{a}\omega_{(n)}^{b}\omega_{(p)}+\frac{1}{2}\left(\omega_{(m)}^{a}\omega_{(n)}^{b}-\Lambda\,e_{(m)}^{a}e_{(n)}^{b}\right)\tau_{(p)}\right) \,,
\end{aligned}
\end{equation}
where the different terms can be rearranged in terms of $\mu_{(n)}$ as
\begin{equation}\label{csactiongalinf3}
S_{\text{CS}}= \sum_{i=0}^{\infty} \mu_{(i)}  S_{(i)}\,.
\end{equation}
We shall now discuss in more detail the first three terms in the sum.
\begin{itemize}
\item The $\mu_{(0)}$ term can be directly read off from the second sum in \eqref{csactiongalinf2}. Up to total derivatives it reads
\begin{equation}\label{lagrangian0}
S_{(0)} =-2\int \tau_{(0)}d\omega_{(0)}\,.
\end{equation}
which corresponds to Galilean gravity in 2+1 dimensions \cite{Bergshoeff:2017btm}.
\item In order to write down the $\mu_{(1)}$ term in a familiar way, we first note that we can relate the curvatures associated to the different generators
\eqref{curvatureform2} in the form
\begin{equation}
\begin{aligned}
\omega^{a}_{(0)} F_a\left[ P^{(0)} \right] &=  e^{a}_{(0)} F_a\left[ G^{(0)} \right]-\epsilon_{ab}\left(\omega_{(0)}^{a} \omega_{(0)}^{b} +\Lambda e_{(0)}^{a} e_{(0)}^{b} \right) \tau_{(0)} + {\rm T.D.} \,, \\
\omega_{(1)}F\left[H^{(0)}\right] &= \tau_{(0)}F\left[J^{(1)}\right] -\frac{1}{2}\epsilon_{ab}\left(\omega_{(0)}^{a} \omega_{(0)}^{b} -\Lambda e_{(0)}^{a} e_{(0)}^{b}\right)\tau_{(0)} + {\rm T.D.} \,, \\
\omega_{(0)}F\left[H^{(1)}\right] & = \tau_{(1)}F\left[J^{(0)}\right] +\epsilon_{ab} e_{(0)}^{a} \omega_{(0)}^{b}\omega_{(0)} + {\rm T.D.} \,,
\end{aligned}
\end{equation}
where ${\rm T.D.}$ stands for total derivative terms. Using these relations, the action $S_{(1)}$ takes the form
\begin{equation}\label{lagrangian1-2}
S_{(1)}= 2\int \Bigg( e^{a}_{(0)} F_a\left[ G^{(0)} \right]
-\tau_{(0)}F\left[J^{(1)}\right]-\tau_{(1)}F\left[J^{(0)}\right]- \Lambda\epsilon_{ab} e_{(0)}^{a} e_{(0)}^{b}\tau_{(0)} \Bigg)\,.\\
\end{equation}
This action defines extended Newton--Hooke gravity \cite{Papageorgiou:2010ud,Hartong:2016yrf} and generalises extended Bargmann gravity 
\cite{Papageorgiou:2009zc,Bergshoeff:2016lwr} to include the cosmological constant. This can be seen clearly by relabeling the gauge fields as
\begin{equation}\label{renamefieldsL1}
\begin{aligned}
e^a_{(0)}&=e^a\,,& 
\tau_{(0)}&=\tau \,,& 
\tau_{(1)}&=m \,,&\\
\omega^a_{(0)}&=\omega^a\,,&
\omega_{(0)}&=\omega\,,&
\omega_{(1)}&=s\,.
\end{aligned}
\end{equation}

\item For the $\mu_{(2)}$ term we use the relations
\begin{equation}
\begin{aligned}
\omega^{a}_{(0)} F_a\left[ P^{(1)} \right] &=  e^{a}_{(1)} F_a\left[ G^{(0)} \right]-\epsilon_{ab} \left(
\omega_{(0)}^{a} \omega_{(0)}^{b}\tau_{(1)} +
\omega_{(0)}^{a} \omega_{(1)}^{b}\tau_{(0)} +
\omega_{(0)}^{a} e_{(0)}^{b}\omega_{(1)} 
\right)+ {\rm T.D} \,, \\
\omega^{a}_{(1)} F_a\left[ P^{(0)} \right] &=  e^{a}_{(0)} F_a\left[ G^{(1)} \right]-\epsilon_{ab} \left(
\omega_{(1)}^{a} \omega_{(0)}^{b}\tau_{(0)} -
\omega_{(0)}^{a} e_{(0)}^{b}\omega_{(1)} 
\right)+ {\rm T.D.} \,, \\
\omega_{(0)}F\left[H^{(2)}\right] & = \tau_{(2)}F\left[J^{(0)}\right] +\epsilon_{ab} \left(
e_{(0)}^{a} \omega_{(1)}^{b}\omega_{(0)} +
e_{(1)}^{a} \omega_{(0)}^{b}\omega_{(0)}
\right)+ {\rm T.D.} \,,\\
\omega_{(1)}F\left[H^{(1)}\right] & = \tau_{(1)}F\left[J^{(1)}\right] +\epsilon_{ab} \left(
e_{(0)}^{a} \omega_{(0)}^{b}\omega_{(1)} -
\frac{1}{2}\omega_{(0)}^{a} \omega_{(0)}^{b}\tau_{(1)}
\right)+ {\rm T.D.} \,,\\
\omega_{(2)}F\left[H^{(0)}\right] & = \tau_{(0)}F\left[J^{(2)}\right] -\epsilon_{ab} \omega_{(0)}^{a} \omega_{(1)}^{b}\tau_{(0)} + {\rm T.D.} \,,
\end{aligned}
\end{equation}
which allows one to express $S_{(2)}$ as
\begin{equation}\label{lagrangian2-2}
\begin{aligned}
S_{(2)} &= 2\int \Bigg( e^{a}_{(0)} F_a\left[ G^{(1)} \right]+e^{a}_{(1)} F_a\left[ G^{(0)} \right]\\
&\quad\quad\qquad-\tau_{(0)}F\left[J^{(2)}\right]  -  \tau_{(1)}F\left[J^{(1)}\right]  -  \tau_{(2)}F\left[J^{(0)}\right] \\
&\quad\quad\qquad-\Lambda\ \epsilon_{ab} \left( e^{a}_{(0)} e^{b}_{(0)}\tau_{(1)} +e^{a}_{(0)} e^{b}_{(1)}\tau_{(0)}+e^{a}_{(1)}e^{b}_{(0)}\tau_{(0)}\right) \Bigg) \,.
\end{aligned}
\end{equation}
For $\Lambda=0$, this is precisely the Lagrangian for non-relativistic three-dimensional gravity found in \cite{Ozdemir:2019orp}, where the gauge fields should identified as in \eqref{renamefieldsL2} together with
\begin{equation}\label{renamefieldsL2}
\begin{aligned}
e^a_{(1)}&=t^a\,,& 
\tau_{(2)}&=y \,,
\\
\omega^a_{(1)}&=b^a\,,&
\omega_{(2)}&=z\,. 
\end{aligned}
\end{equation}

\end{itemize}
In general the action $S_{(i)}$ can be written, after partial integration, in the form
\begin{equation}
\begin{aligned}
S_{(i)}=2\int&\Bigg( \sum_{m,n=0}^{\infty}
\left(
 \delta_{i}^{m+n+1}\,e^a_{(m)}F_a\left[G^{(n)}\right]+ \delta_{i}^{m+n}\,\tau_{(m)}F\left[J^{(n)}\right]\right) \\
&\qquad\qquad-\Lambda \sum_{m,n,p=0}^{\infty} \delta_{i}^{m+n+p+1}\,\epsilon_{ab}e^a_{(m)}e^b_{(n)}\tau_{(p)}\Bigg)
\end{aligned}
\end{equation}
It is important to note that the infinite sum \eqref{csactiongalinf3} can be consistently truncated to give
\begin{equation}\label{condND}
S_{\text{CS}}= \sum_{i=0}^{i_0} \mu_{(i)} S_i\,,
\end{equation}
which gives an action for the gauge fields
\begin{align}
e^a_{(m)},\; \omega^a_{(m)},\quad&{\rm for}\;\;m=0,1,\dots,i_0\,,& \tau_{(n)},\;\omega_{(n)},\quad&{\rm for}\;\;n=0,1,\dots,i_0+1\,,
\end{align}
and sets all the other gauge fields to zero. In this case, the invariant tensor \eqref{invtensor} in non-degenerate for $\mu_{(i_0+1)}\neq 0$, and the truncation leads to the condition
\begin{equation}
\mu_{(m)}=0 \,,\quad {\rm for} \quad m>i_0+1 \,.
\end{equation}
This corresponds to the Chern--Simons action for non-relativistic gravity invariant under the expanded algebra \eqref{redresp} for $N=2(i_0+1)$,
 \begin{equation}
\Big(s_{0}^{(2( i_0+1))}\times\{\tilde H, \tilde J\}
\Big)\oplus\Big(s_{1}^{(2( i_0+1))}\times\{\tilde P_{a}, \tilde G_{b}\} \Big)\,.
 \end{equation}
This means that only the expanded algebras with even values of $N$ give rise to Chern--Simons actions with a non-degenerate invariant bilinear form.

\subsubsection*{AdS case}
Let us consider now the general non-relativistic algebra \eqref{galkmalg2+1} in the case of negative cosmological constant and set $\Lambda=-\ell^{-2}<0$. In this case it takes the form
\begin{multicols}{2}
\begin{subequations}\label{nhads2+1}
\setlength{\abovedisplayskip}{-15pt}
\allowdisplaybreaks
\begin{align}
[J^{(m)},P_{a}^{(n)}] &=-\epsilon_{a}^{\;\;b}P_{b}^{(m+n)}\,,\\
[G_{a}^{(m)},G_{b}^{(n)}] &=\epsilon_{ab}J^{(m+n+1)}\,,\\
[J^{(m)},G_{a}^{(n)}] &=-\epsilon_{a}^{\;\;b}G_{b}^{(m+n)}\,,\\
[G_{a}^{(m)},P_{b}^{(n)}] &=\epsilon_{ab}H^{(m+n+1)}\,,\\
[H^{(m)},G_{a}^{(n)}] &=-\epsilon_{a}^{\;\;b}P_{b}^{(m+n)}\,,\\
[H^{(m)},P_{a}^{(n)}] &=-\frac{1}{\ell^2} \epsilon_{a}^{\;\;b}G_{b}^{(m+n)}\,,\\
[P_{a}^{(m)},P_{b}^{(n)}] &=\frac{1}{\ell^2}\epsilon_{ab}J^{(m+n+1)}\,.
\end{align}
\end{subequations}
\end{multicols}
\noindent 
By defining the change of basis
\begin{align}\label{chiralbasisExpNH}
L_{a}^{\pm(m)}&=\frac{1}{2}\left(  G_{a}^{(m)} \pm \ell P_{a}^{(m)}\right) \,,&
L^{\pm(m)}&=\frac{1}{2}\left( J^{(m)} \pm \ell H^{(m)} \right)\,,
\end{align}
this algebra can be written as the direct sum $\left\{ L_{a}^{+(m)}, L^{+(m)} \right\} \oplus \left\{ L_{a}^{-(m)}, L^{-(m)} \right\}$, where 
\begin{align}
[  L^{\pm (m)},   L^{\pm(n)}_{a}]&=-\epsilon_{a}^{\;\;b}   L^{\pm(m+n)}_{b}\, ,&
[  L^{\pm(m)}_{a},   L^{\pm(n)}_{b}]&=  \epsilon_{ab}   L^{\pm(m+n+1)}\,.
\end{align}
The invariant tensor in this basis follows from \eqref{invtensor} and has takes the form
\begin{align}\label{invtensorsplit}
\left\langle L_{a}^{\pm(m)}L_{b}^{\pm(n)}\right\rangle &=\pm\frac{\ell}{2}\mu_{(m+n+1)}\delta_{ab}\,,&\left\langle L^{\pm(m)}L^{\pm(n)}\right\rangle &=\mp \frac{\ell}{2}\mu_{(m+n)}\,.
\end{align}
This isomorphism implies that one can reformulate the expanded Newton--Hooke gravity action \eqref{csactiongalinf2} in the form
\begin{equation}\label{cs-cs}
S_{\text{CS}}[A]=S_{\text{CS}}[A^+]-S_{\text{CS}}[A^-]\,,
\end{equation}
where the gauge connections are given by
\begin{equation}\label{chiralconnections}
A^\pm = \sum_{m=0}^\infty\left[ \left( \omega^a_{(m)} \pm \frac{1}{\ell} e^a_{(m)}\right) L^{(m)}_a+ \left( \omega_{(m)} \pm \frac{1}{\ell} \tau_{(m)} \right) L^{(m)}\right]\,,
\end{equation}
and the $\left\{ L^{(m)}_a,L^{(m)}\right\}$ is now a single set of generators satisfying
\begin{align}\label{NRlorentz}
[  L^{(m)},   L^{(n)}_{a}]&=-\epsilon_{a}^{\;\;b}   L^{(m+n)}_{b}\, ,&
[  L^{(m)}_{a},   L^{(n)}_{b}]&=  \epsilon_{ab}   L^{(m+n+1)}\,,\\
\left\langle L_{a}^{(m)}L_{b}^{(n)}\right\rangle &=\frac{1}{2}\gamma_{(m+n+1)}\delta_{ab}\,,&\left\langle L^{(m)}L^{(n)}\right\rangle &= \frac{1}{2}\gamma_{(m+n)}\,.
\end{align}
This is in completely analogy with the relativistic case, where three-dimensional Einstein gravity with negative cosmological constant can be reformulated as a Chern--Simons theory of the form \eqref{cs-cs} with chiral connections taking values in the $\mathfrak{sl}(2,\mathbb{R})$ algebra \cite{Achucarro:1987vz,Witten:1988hc,Coussaert:1995zp,Banados:1998gg,Carlip:2005zn}. In fact, the algebra \eqref{NRlorentz} can be obtained as a non-relativistic $S^{(\infty)}$ expansion of $\mathfrak{so}(2,1)\approx \mathfrak{sl}(2,\mathbb{R})$ by defining
\begin{equation}
L^{(m)}=\lambda_{2m} \otimes \tilde L \,,\qquad L_a^{(m)}=\lambda_{2m+1}  \otimes \tilde L_a\,,
\end{equation}
where $\tilde L_A=\{\tilde L_0 \equiv \tilde L , \tilde L_a \}$ are the generators of the Lorentz algebra in 2+1 dimensions and satisfy $\left[\tilde L_A, \tilde L_B \right]=\epsilon^C_{\;\;AB}\tilde L_C$. Once again, quotients by suitable ideals reproduce $S_E^{(N)}$ expansions. The case $N=1$ shows that the Newton--Hooke algebra in 2+1 dimensions is isomorphic to two copies of the Euclidean algebra in 1+1 dimensions, while the $N=2$ case expresses the extended Newton--Hooke symmetry in three-dimensions as two copies of the centrally extended Newton-Hooke symmetry in two dimensions \cite{Alvarez:2007fw}, isomorphic to the Nappi--Witten algebra \cite{Hartong:2017bwq,Penafiel:2019czp}.

\subsubsection*{Exotic invariant bilinear form}

As $\mathfrak{so}(2,2)$ is semi-simple there are two independent invariant bilinear forms. Besides~\eqref{eq:so22inv1}, a second invariant bilinear form on $\mathfrak{so}(2,2)$ is 
\begin{align}\label{exoticinvtads}
\left\langle J_{AB} J_{CD} \right\rangle &= \eta_{AD}\eta_{BC}-\eta_{AC}\eta_{BD}\,,& 
\left\langle P_{A} P_{B} \right\rangle &= -\Lambda\eta_{AB}\,,
\end{align}
which induces the following invariant tensor for the expanded generators \eqref{expgenkm}
\begin{align}
\left\langle  J_{ab}^{(m)}   J_{cb}^{(n)}  \right\rangle &=\beta_{(2m+2n)}\left(\delta_{ad}\delta_{bc}-\delta_{ac}\delta_{bd}\right)\,,
&\left\langle  B_{a}^{(m)}   B_{b}^{(n)}  \right\rangle &=\beta_{(2m+2n+2)}\delta_{ab}\,,\\
\left\langle  H^{(m)}   H^{(n)}  \right\rangle &=\Lambda \beta_{(2m+2n)}\,,
&\left\langle  P_{a}^{(m)}   P_{b}^{(n)}  \right\rangle &=-\Lambda \beta_{(2m+2n+2)}\delta_{ab}\,.
\end{align}
Defining $\beta_{(2m)}=\nu_{(m)}$ and using \eqref{redefJP} this leads to
\begin{align}
\left\langle  J^{(m)}   J^{(n)}  \right\rangle &=-\nu_{(m+n)}\,,
&\left\langle  G_{a}^{(m)}   G_{b}^{(n)}  \right\rangle &=\nu_{(m+n+1)}\delta_{ab}\,,\\
\left\langle  H^{(m)}   H^{(n)}  \right\rangle &=\Lambda \nu_{(m+n)}\,,
&\left\langle  P_{a}^{(m)}   P_{b}^{(n)}  \right\rangle &=-\Lambda \nu_{(m+n+1)}\delta_{ab}\,.\label{exoticpairing2+1}
\end{align}
In this case, the action \eqref{csactiongalinf} boils down to
\begin{equation}\label{csactiongalinf2-2}
\begin{aligned}
S_{\text{CS}} =&\sum_{m,n=0}^{\infty}\nu_{(m+n+1)}\int \Bigg( \omega_{(m)}^{a}F_a\left[G^{(n)}\right]-\Lambda e_{(m)}^{a}F_a\left[P^{(n)}\right]\Bigg)\\
-&\sum_{m,n=0}^{\infty}\nu_{(m+n)}\int \Bigg(\omega_{(m)}F\left[J^{(n)}\right]-\Lambda \tau_{(p)}F\left[H^{(q)}\right]\Bigg)\\
+&\sum_{m,n,p=0}^{\infty}\nu_{(m+n+p+1)}\,\int \epsilon_{ab}\left(-\Lambda e_{(m)}^{a}\omega_{(n)}^{b}\tau_{(p)}+\frac{1}{2}\left(\omega_{(m)}^{a}\omega_{(n)}^{b}-\Lambda\,e_{(m)}^{a}e_{(n)}^{b}\right)\omega_{(p)}\right)\,.
\end{aligned}
\end{equation}
The exotic invariant tensor in the chiral basis \eqref{chiralbasisExpNH} takes the form
\begin{align}\label{invtensorsplitexotic}
\left\langle L_{a}^{\pm(m)}L_{b}^{\pm(n)}\right\rangle &=\frac{1}{2}\nu_{(m+n+1)}\delta_{ab}\,,&\left\langle L^{(m)}L^{(n)}\right\rangle &=-\frac{1}{2}\nu_{(m+n)}\,.
\end{align}
Therefore, the action \eqref{csactiongalinf2-2} can be reformulated in terms of \eqref{chiralconnections} as
\begin{equation}
S_{\text{CS}}[A]=S_{\text{CS}}[A^+]+S_{\text{CS}}[A^-]\,,
\end{equation}
where now the relative sign between the two Chern--Simons actions is changed compared to \eqref{cs-cs}.

\subsection{Extended Carrollian gravities}

In the Carrollian case, the definition of the expanded generators with algebra given in~\eqref{carrollkmalg} together with \eqref{expandedinvtensor} yields the following invariant tensor 
\begin{align}
\left\langle B_{a}^{(m)}P_{b}^{(n)}\right\rangle &=-\alpha_{(2m+2n+1)}\epsilon_{ab}\,, & 
\left\langle J_{ab}^{(m)}H^{(n)}\right\rangle &=-\alpha_{(2m+2n+1)}\epsilon_{ab}\,.  \label{invtensor0carroll}
\end{align}
Thus, in contrast with the Newton--Hooke case, the constants $\alpha_{(\gamma)}$ in \eqref{invtensor0} are non-vanishing only for odd values of $\gamma$, and we will relabel them as $\rho_{(m)}=\alpha_{(2m+1)}$. Using \eqref{expgencarrollkm} and the dual generators \eqref{redefJP} the invariant bilinear form in the $D=3$ case can also be written as
\begin{align}\label{invtensorcarroll}
\left\langle G_{a}^{(m)}P_{b}^{(n)}\right\rangle &=\rho_{(m+n)}\delta_{ab}\,,&\left\langle J^{(m)}H^{(n)}\right\rangle &=-\rho_{(m+n)}\,.
\end{align}
The infinite-dimensional algebra \eqref{carrollkmalg} takes the form
\begin{multicols}{2}
\begin{subequations}\label{gcarrollkmalg2+1}
\setlength{\abovedisplayskip}{-15pt}
\allowdisplaybreaks
\begin{align}
[J^{(m)},P_{a}^{(n)}] &=-\epsilon_{a}^{\;\;b}P_{b}^{(m+n)}\,,\\
[G_{a}^{(m)},G_{b}^{(n)}] &=\epsilon_{ab}J^{(m+n+1)}\,,\\
[J^{(m)},G_{a}^{(n)}] &=-\epsilon_{a}^{\;\;b}G_{b}^{(m+n)}\,,\\
[G_{a}^{(m)},P_{b}^{(n)}] &=\epsilon_{ab}H^{(m+n)}\,,\\
[H^{(m)},G_{a}^{(n)}] &=-\epsilon_{a}^{\;\;b}P_{b}^{(m+n+1)}\,,\\
[H^{(m)},P_{a}^{(n)}] &=\Lambda \epsilon_{a}^{\;\;b}G_{b}^{(m+n)}\,,\\
[P_{a}^{(m)},P_{b}^{(n)}] &=-\Lambda\epsilon_{ab}J^{(m+n)}\,.
\end{align}
\end{subequations}
\end{multicols}\noindent

\subsubsection*{Action}
In the Carrollian case, the connection one-form \eqref{csconnection} together with the commutation relations \eqref{gcarrollkmalg2+1} yields a curvature two-form \eqref{curvatureform} with the following components:
\begin{equation}\label{curvatureformcarroll2}
\begin{aligned}
F^{a}\left[P^{(m)}\right] & =de_{(m)}^{a}
-\sum_{n,p=0}^{\infty}\epsilon_{\;b}^{a}
\left(\delta_{m}^{n+p+1}\,
\omega_{(n)}^{b}\tau_{(p)}
-\delta_{m}^{(n+p)} \, e_{(n)}^{b}\omega_{(p)}\right) \\
F^{a}\left[G^{(m)}\right] & =d\omega_{(m)}^{a}-\sum_{n,p=0}^{\infty}\delta_{m}^{n+p}\,\epsilon_{\;b}^{a}\; \left(\omega_{(n)}^{b}\omega_{(p)}-\Lambda e_{(p)}^{b}\tau_{(n)}\right) \,,\\
F\left[H^{(m)}\right] & =d\tau_{(m)}+\sum_{n,p=0}^{\infty}\delta_{m}^{n+p}\,\epsilon_{ab}\;e_{(p)}^{a}\omega_{(n)}^{b}\,,\\
F\left[J^{(m)}\right] & =d\omega_{(m)}+\frac{1}{2}\sum_{n,p=0}^{\infty} \epsilon_{ab}
\left(
\delta_{m}^{n+p+1}\,\omega_{(n)}^{a}\omega_{(p)}^{b}-\Lambda\delta_{m}^{n+p}\,e_{(n)}^{a}e_{(p)}^{b}\right)\,.\\
\end{aligned}
\end{equation}
The Chern--Simons action \eqref{csactiongalinf} then takes the form
\begin{equation}\label{csactioncarinf2}
\begin{aligned}
S_{\text{CS}} &=\sum_{m,n=0}^{\infty}\rho_{(m+n)}\int \Bigg( e_{(m)}^{a}F_a\left[G^{(n)}\right]+\omega_{(m)}^{a}F_a\left[P^{(n)}\right]
-\tau_{(m)}F\left[J^{(n)}\right]-\omega_{(m)}F\left[H^{(n)}\right]
\Bigg)\\
&\quad +\sum_{m,n,p=0}^{\infty}\int \epsilon_{ab}
\Bigg(\rho_{(m+n+p)}\left(
e_{(m)}^{a}\omega_{(n)}^{b}\omega_{(p)}-\frac{\Lambda}{2}\,e_{(m)}^{a}e_{(n)}^{b}\tau_{(p)}\right)
+\frac{1}{2}\rho_{(m+n+p+1)}\,\omega_{(m)}^{a}\omega_{(n)}^{b}\tau_{(p)}\Bigg)\,,
\end{aligned}
\end{equation}
which can be also written as
\begin{equation}\label{csactioncarinf3}
S_{\text{CS}}= \sum_{i=0}^{\infty} \rho_{(i)}  S_{(i)}\,,
\end{equation}
where after integration by parts we can write $S_{(i)}$ as
\begin{equation}
S_{(i)}=2\int\Bigg(\sum_{m,n=0}^{\infty}  \delta_{i}^{m+n}\,\left( e^a_{(m)}F_a\left[G^{(n)}\right]-\tau_{(m)}F\left[J^{(n)}\right]\right)- \Lambda \sum_{m,n,p=0}^{\infty} \delta_{i}^{m+n+p}\,\epsilon_{ab}e^a_{(m)}e^b_{(n)}\tau_{(p)}\Bigg)\,.
\end{equation}
The first action in this sequence is given by
\begin{equation}
S_{(0)}=2\int\left(
e^a F_a\left[G\right]-\tau F\left[J\right]- \Lambda \epsilon_{ab} e^a e^b\,\tau \right) \,,
\end{equation}
where we have used the field redefinition \eqref{renamefieldsL1}. This corresponds to Carrollian AdS Chern-Simons gravity \cite{Ali:2019jjp,Ravera:2019ize} and reduces to three-dimensional Carroll gravity \cite{Hartong:2015xda,Bergshoeff:2016soe,Matulich:2019cdo} in the vanishing cosmological constant limit.\footnote{The first realisations of the Carroll symmetry in gravity appeared in the study of the zero signature limit
\cite{Henneaux:1979vn,Teitelboim:1981ua} or strong coupling limit \cite{Isham:1975ur,Henneaux:1981su} (see also \cite{Anderson:2002zn,Niedermaier:2014xwa}) of general relativity, the last one being closely related to the Belinski--Khalatnikov--Lifshitz (BKL) limit of gravity \cite{Belinsky:1970ew,Belinsky:1982pk,Damour:2002et}. On the other hand, the ultra-relativistic expansion of general relativity has been explored in \cite{Dautcourt:1997hb}, while the Carrollian limit at the level of the Einstein--Hilbert action has been studied in \cite{Bergshoeff:2017btm}. It is also important to mention the realisation of the Carroll symmetry in the near horizon limit of black holes \cite{Donnay:2019jiz}.
}

Using also \eqref{renamefieldsL2}, the second term of the sum reads
\begin{equation}
S_{(1)}=2\int\bigg(
e^a F_a\left[B\right]+t^a F_a\left[G\right]-\tau F\left[S\right]-m F\left[J\right]
- \Lambda \epsilon_{ab} \left( e^a e^b\,m+2e^a t^b\,\tau\right)\bigg)
\end{equation}
and defines a Carrollian counterpart of (post-)Newtonian gravity action \cite{Hansen:2018ofj} in three dimensions including cosmological constant. Similarly, the next actions in the sequence define further post-Carrollian corrections for three-dimensional Carroll (A)dS gravity.

In is interesting that unlike the Newton--Hooke expansion, truncations of the general Carrollian action \eqref{csactioncarinf3} for arbitrary $r$ lead to a non-degenerate invariant bilinear form for the corresponding truncation in \eqref{gcarrollkmalg2+1}. Also, the extended Carrollian algebras for (A)dS do not include central terms, which is in contrast with the Newton--Hooke case, where the central extensions are precisely the ones that allow us to find non-generate pairings when \eqref{condND} holds.

\subsubsection*{Exotic invariant bilinear form}

We can also consider the exotic invariant tensor for (A)dS \eqref{exoticinvtads}, which leads to the following expression for pairings of the expanded generators \eqref{expgencarrollkm}
\begin{align}
\left\langle  J_{ab}^{(m)}   J_{cb}^{(n)}  \right\rangle &=\beta_{(2m+2n)}\left(\delta_{ad}\delta_{bc}-\delta_{ac}\delta_{bd}\right)\,,
&\left\langle  B_{a}^{(m)}   B_{b}^{(n)}  \right\rangle &=\beta_{(2m+2n+2)}\delta_{ab}\,,\\
\left\langle  H^{(m)}   H^{(n)}  \right\rangle &=\Lambda \beta_{(2m+2m+2)}\,,
&\left\langle  P_{a}^{(m)}   P_{b}^{(n)}  \right\rangle &=-\Lambda \beta_{(2m+2n)}\delta_{ab}\,.
\end{align}
Defining $\beta_{(2m)}=\sigma_{(m)}$ and using \eqref{redefJP} this can also be written as
\begin{subequations}
\begin{align}
\left\langle  J^{(m)}   J^{(n)}  \right\rangle &=-\sigma_{(m+n)}\,,
&\left\langle  G_{a}^{(m)}   G_{b}^{(n)}  \right\rangle &=\sigma_{(m+n+1)}\delta_{ab}\,,\\
\left\langle  H^{(m)}   H^{(n)}  \right\rangle &=\Lambda \sigma_{(m+n+1)}\,,
&\left\langle  P_{a}^{(m)}   P_{b}^{(n)}  \right\rangle &=-\Lambda \sigma_{(m+n)}\delta_{ab}\,.\label{expexoticpairing2+1C}
\end{align}
\end{subequations}
In this case, the action \eqref{csactiongalinf} reduces to
\begin{equation}\label{csactioncarinf2-2}
\begin{aligned}
 S_{\text{CS}} =&\sum_{m,n=0}^{\infty}\int \Bigg(
 \sigma_{(m+n+1)}\left(
  \omega_{(m)}^{a}F_a\left[G^{(n)}\right]-\Lambda \tau_{(m)}F\left[H^{(n)}\right]
  \right)\\
&\hspace{30mm} -\sigma_{(m+n)} \left(\omega_{(m)}F\left[J^{(n)}\right]+\Lambda e_{(m)}^{a}F_a\left[P^{(n)}\right]\right)
 \Bigg)
\\
+&\sum_{m,n,p=0}^{\infty}\epsilon_{ab} \int \Bigg( \, \sigma_{(m+n+p+1)}\left[-\Lambda e_{(m)}^{a}\omega_{(n)}^{b}\tau_{(p)}+\frac{1}{2}\omega_{(m)}^{a}\omega_{(n)}^{b}\omega_{(p)}\right]
\\
&\hspace{60mm}-\frac{\Lambda}{2}\sigma_{(m+n+p)}\,e_{(m)}^{a}e_{(n)}^{b}\omega_{(p)}\Bigg)
\,.
\end{aligned}
\end{equation}
One can consider again finite truncations of this actions and, which produce exotic Carroll and post-Carrollian gravity theories in $2+1$ dimensions.

\section{Conclusions and Outlook}

We have studied the non-relativistic symmetries obtained by Lie algebra expansion of the AdS or dS algebra in $D$ space-time dimensions, i.e.  $\mathfrak{so}(D-1,2)$ or $\mathfrak{so}(D,1)$. This generalises previous constructions to include a cosmological constant and generates an infinite family of algebras of Newton--Hooke \cite{Bacry:1968zf,derome1972hooke} or Carrollian type. Subsequently we have shown how these infinite-dimensional symmetries can be embedded in in different Kac-Moody algebras. Moreover, the different finite $S_E^{(N)}$ expansions contained in each family can be systematically constructed as a free algebra and correspond to suitable quotients of the infinite-dimensional case.

We have considered a gravitational model based on them by focussing on the case of $(2+1)$-dimensional Chern--Simons theories. We show that the family of algebras generates systematically non-relativistic gravity theories extended by a cosmological constant, such as extended Bargmann gravity~\cite{Bergshoeff:2016lwr,Hartong:2016yrf} and (post-)Newtonian gravity~\cite{Ozdemir:2019orp}. 

As stated in \cite{Hansen:2019vqf} these extended Newton--Hooke symmetries encode the large $c$ expansion of general relativity. An interesting way to see this in 2+1 dimensions is that the Chern--Simons formulation of AdS three-dimensional gravity allows to construct the metric out of the chiral connections \eqref{chiralconnections} in the form \cite{Grumiller:2016pqb}
 \begin{equation}\label{metricfromA2}
ds^2=g_{\mu\nu}dx^\mu dx^\nu=\frac{\ell^2}{2}\left\langle \left(A^+ -A^-\right)\otimes \left(A^+ -A^-\right) \right\rangle \,.
 \end{equation}
Using \eqref{chiralconnections} and rearranging the sum in terms of $\gamma_{(n)}$ we get
 \begin{equation}\label{summetric}
\begin{aligned}
ds^2&= - \sum_{m,n=0}^\infty \gamma_{(m+n)}\tau_{(m)}\tau_{(n)}+  \sum_{m,n=0}^\infty\gamma_{(m+n+1)} e^a_{(m)}e^b_{(n)}\,.\\
\end{aligned}
 \end{equation}
The flat version of this metric has been studied in \cite{Gomis:2019sqv} as an infinite extension of Minkowski space where post-Newtonian corrections to relativistic symmetries can be naturally implemented. This metric can be found in the vanishing cosmological constant limit by defining coordinates $x^\mu_{(m)}$ such that $\tau_{(m)}=\tau_{(m)\mu}dx^\mu_{(m)}$ and $e^a_{(m)}=e^a_{(m)\mu}dx^\mu_{(m)}$. Then, the Minkowskian case is obtained by setting $e^a_{(m)\mu}=\delta^a_\mu$ and $\tau_{(m)\mu}=(1,0,0,0)$ for all $m$, which leads to the expanded metric constructed in \cite{Gomis:2019sqv}.

On the other hand, keeping the gauge field components $\tau_{(m)\mu}$ and $e^a_{(m)\mu}$ arbitrary while identifying the coordinates as $x^\mu_{(m)}=x^\mu$, allows one to find the (2+1)-dimensional version of the post-Newtonian expansion of the metric used in \cite{dautcourt1990newtonian,DePietri:1994je,Dautcourt:1996pm,VandenBleeken:2017rij,Hansen:2019vqf,Hansen:2019svu}. Indeed, identifying the invariant tensor constants as powers of the speed of light in the form $\gamma_{(n)}\rightarrow c^{2(1-n)}$, and defining the spatial metrics $h_{\mu\nu}dx^\mu dx^\nu=e^a_{(0)} e_{a(0)}$ and $\Phi_{\mu\nu}dx^\mu dx^\nu=e^a_{(0)} e_{a(1)} +e^a_{(1)} e_{a(0)}$, this expression reproduces the aforementioned non-relativistic expansion:
\begin{equation}
g_{\mu\nu} = - c^{2} \,\tau_\mu \tau_\nu +h_{\mu\nu} -\tau_\mu m_\nu-m_\mu \tau_\nu +c^{-2}\left(\Phi_{\mu\nu}-m_\mu m_\nu-\tau_\mu y_\nu-y_\mu\tau_\nu  \right)+O(c^{-4})\,,
\end{equation}
%\begin{align}
%g_{\mu\nu} = - c^{2} \,\tau_\mu \tau_\nu + h_{\mu\nu} -\tau_\mu m_\nu-\tau_\nu m_\mu
%+c^{-2}\left(\Phi_{\mu\nu}-m_\mu m_\nu-\tau_\mu y_\nu-\tau_\nu y_\mu \right)+O(c^{-4})\,.
%\end{align}
where we have also relabeled the metric fields according to \eqref{renamefieldsL1} and \eqref{renamefieldsL2}. Alternatively one can introduce an expansion of the dreibein forms 
\begin{align}
E^0&=\sum_{m=0}^\infty \lambda_{2m}\,\tau^{(m)}\,, & \quad E^a&=\sum_{m=0}^\infty \lambda_{2m+1}\,e^a_{(m)}\,,
\end{align}
similar to what was done in \cite{Bergshoeff:2019ctr}. Then, by identifying the tangent space metric with the exotic pairing for the translations \eqref{exoticinvtads}, one can write \cite{Campoleoni:2011tn}
\begin{align}
g_{\mu\nu}=
\eta_{AB}E^A E^B= \ell^2  \left\langle E^A P_A \, E^B P_B \right\rangle\,,
\end{align}
which also leads to the expansion of the metric \eqref{summetric} with $\gamma_{(n)}=\nu_{(n)}$, where $\nu_{(n)}$ appears in the pairing of the momentum generators in~\eqref{exoticpairing2+1}.

We have also considered the Carrollian case with cosmological constant to construct a corresponding Chern--Simons theory based on the algebra~\eqref{carrollkmalg}. This reproduces known ultra-relativistic gravity theories in $2+1$ dimensions and an infinite family of generalisations.

Applying the same procedure as~\eqref{summetric} in the the case of expanded Carrollian (A)dS algebras \eqref{gcarrollkmalg2+1} leads to a novel expansion for the metric tensor that reads
 \begin{equation}\label{Carrollsummetric}
\begin{aligned}
g_{\mu\nu}&= - \sum_{m,n=0}^\infty \sigma_{(m+n+1)}\tau_{(m)\mu}\tau_{(n)\nu}+  \sum_{m,n=0}^\infty\sigma_{(m+n)} e^a_{(m)\mu}e^b_{(n)\nu}\\
&=  \sigma_{(0)}h_{\mu\nu}+ \sigma_{(1)}\left(\Phi_{\mu\nu} 
- \tau_{\mu} \tau_{\nu} \right)
+\sigma_{(2)}\left(\Psi_{\mu\nu} -\tau_\mu m_\nu- m_\mu \tau_\nu
\right)+\dots\,.
\end{aligned}
 \end{equation}
where we have also defined $\Psi_{\mu\nu}dx^\mu dx^\nu=e^a_{(1)} e_{a(1)}+e^a_{(0)} e_{a(2)}+e^a_{(2)} e_{a(0)}$. This metric can be conjectured to describe the ultra-relativistic expansion of three-dimensional Einstein gravity. To evaluate whether this expansion can be generalised to higher dimensions in the context of the Carroll limit of general relativity along the lines of \cite{Henneaux:1979vn} or \cite{Dautcourt:1997hb} is an interesting question that we hope to address in the future.

There are several directions in which this research could be extended. One would be
to generalise our work of symmetries of post-Newtonian correction in the flat Minkowski space~\cite{Gomis:2019sqv} to curved (A)dS
space and construct particle actions in the curved generalisation of the infinite-dimensional Minkowski space. On the other hand it would be interesting to analyse particle systems in the presence of a constant background field by considering the non-relativistic expansions of the Maxwell algebra according to Appendix \ref{sec:maxwell}. One could also
construct actions for extended objects by means of the $p$-brane symmetries outlined in Appendix \ref{sec:brane}.
Finally, it would be interesting to extend the analysis to the supersymmetric (A)dS case.

\subsection*{Acknowledgements}
We acknowledge discussions with Luis Avil\'es, Eric Bergshoeff, Diego Hidalgo, Diederik Roest, Patricio Salgado and Jorge Zanelli. JG acknowledges the hospitality and support of the Van Swinderen Institute. JG also has been supported in part by MINECO FPA2016-76005-C2-1-P and Consolider CPAN, and by the Spanish government (MINECO/FEDER) under project MDM-2014-0369 of ICCUB (Unidad de Excelencia Mar\`a de Maeztu). {PS-R} acknowledges DI-VRIEA for financial support through Proyecto Postdoctorado 2019 VRIEA-PUCV.

\appendix

\section{Further generalisations}

The procedure for obtaining non-relativistic expansions of a given relativistic algebra can be generalised to several other interesting cases. Here, we apply the general scheme  to obtain  non-relativistic Maxwell algebras and non-relativistic brane symmetries.

\subsection{Non-relativistic expansions of the Maxwell algebra}
\label{sec:maxwell}
The relativistic Maxwell algebra in $D$ space-time dimensions is given is given by the following extension of the Poincar\'e algebra \cite{Schrader:1972zd} (see also \cite{Bonanos:2008ez,Gomis:2017cmt,Salgado-Rebolledo:2019kft} and references therein)
\begin{multicols}{2}
\begin{subequations}\label{Maxwellalg}
\setlength{\abovedisplayskip}{-15pt}
\allowdisplaybreaks
\begin{align}
[\tilde J_{AB}, \tilde P_{C}] & = 2\eta_{C[B} \tilde P_{A]}\, , \\
[\tilde J_{AB}, \tilde J_{CD}] &= 4\eta_{[A[C }\tilde J_{D]B]}\, ,\\
[\tilde J_{AB}, \tilde Z_{CD}] &= 4\eta_{[A[C }\tilde Z_{D]B]}\, ,\\
[\tilde P_{A}, \tilde P_{B}]& = \tilde Z_{AB}\, .
\end{align}
\end{subequations}
\end{multicols}\noindent
There is no cosmological constant present in this algebra.

As in the (A)dS case, we decompose the relativistic indices in the time and space components, 
$A=(0,a)$, where $a=1,\ldots,D-1$, and relabel the Lie algebra 
generators using \eqref{gendecpoincare} plus the corresponding relation for the new generator $\tilde{Z}_{AB}$ given by
\begin{equation}
\tilde Z_{AB}\rightarrow\{  \tilde Z_{a} \equiv\tilde Z_{0a} \,,\;\tilde Z_{ab}\}\, .
\end{equation}

In the following, we consider the infinite-dimensional
semigroup $S^{(\infty)}$ given in \eqref{sinfty} and derive Galilean and Carrollian expansions of the Maxwell algebra by choosing different resonant subspace decompositions.

\subsubsection*{Galilean expansions}
Galilean expansions of the Maxwell algebra are determined by the following $\mathbb{Z}_{2}$-graded subspace decomposition
\begin{align}
V_{0}&=\{ \tilde J_{ab},\tilde H, \tilde Z_{ab}\}\,,& V_{1}&=\{\tilde G_{a}, \tilde P_{a},\tilde Z_{a}\}\,,
\end{align}
and the resonant non-reduced expansion
 \begin{equation}
\Big(S_{0}^{(\infty)}\times\left\{ \tilde J_{ab},\tilde H, \tilde Z_{ab}\right\} \Big)\oplus\Big(S_{1}^{(\infty)}\times\left\{ \tilde G_{a}, \tilde P_{a},\tilde Z_{a}\right\} \Big)
 \end{equation}
where $S_0^{(\infty)}$ and $S_1^{(\infty)}$ are given in \eqref{infiniteS0S1}. Defining the expanded generators in the form
 \begin{align}
J_{ab}^{(m)}&=\lambda_{2m}\otimes \tilde  J_{ab}\,, & B_{a}^{(m)}&=\lambda_{2m+1}\otimes \tilde  G_{a}\,,\nn\\
H^{(m)}&=\lambda_{2m}\otimes \tilde  H\,,& P_{a}^{(m)}&=\lambda_{2m+1}\otimes \tilde  P_{a}\,,\\
Z_{ab}^{(m)}&=\lambda_{2m}\otimes \tilde Z_{ab}\,, & Z_{a}^{(m)}&=\lambda_{2m+1}\otimes \tilde  Z_{a}\nn \,,
 \end{align}
leads to the following infinite-dimensional algebra
\begin{multicols}{2}
\begin{subequations}\label{Maxwellinfinite}
\setlength{\abovedisplayskip}{-15pt}
\allowdisplaybreaks
\begin{align}
[  B^{(m)}_{a},  H^{(n)}]&=  P^{(m+n)}_{a}\, ,\\
[  B^{(m)}_{a},  P^{(n)}_{b}]&=\delta_{ab}  H^{(m+n+1)}\, , \\
[  J^{(m)}_{ab},  P^{(n)}_{c}]&= 2\delta_{c[b}  P^{(m+n)}_{a]}\, ,\\
[  B^{(m)}_{a},  B^{(n)}_{b}]&=  J^{(m+n+1)}_{ab}\, ,\\
[  J^{(m)}_{ab},  B^{(n)}_{c}]&=2\delta_{c[b}  B^{(m+n)}_{a]}\, ,\\
[  J^{(m)}_{ab},  J^{(n)}_{cd}]&=4\delta_{[a[c}  J^{(m+n)}_{d]b]}\, ,\\
[  J^{(m)}_{ab},  Z^{(n)}_{cd}]&=4\delta_{[a[c}  Z^{(m+n)}_{d]b]}\, ,\\
[  J^{(m)}_{ab},  Z^{(n)}_{c}]&=2\delta_{c[b}  Z^{(m+n)}_{a]}\, ,\\
[  Z^{(m)}_{ab},  B^{(n)}_{c}]&=2\delta_{c[b}  Z^{(m+n)}_{a]}\, ,\\
[  B^{(m)}_{a},  Z^{(n)}_{b}]&=  Z^{(m+n+1)}_{ab}\, ,\\
[  P^{(m)}_{a},  H^{(n)}]&=-  Z^{(m+n)}_{a}\, ,\\
[  P^{(m)}_{a},  P^{(n)}_{b}]&=   Z^{(m+n+1)}_{ab}\, .
\end{align}
\end{subequations}
\end{multicols}\noindent

Truncations of this infinite-dimensional algebra by different ideals lead to expansions with finite semigroups $S_{E}^{(N)}$ \eqref{semigroupsn}. For $N=1$, we get the electric non-relativistic Maxwell algebra \cite{Gomis:2019fdh}, while the $N=2$ case leads to a generalisation of the exotic Maxwellian Bargmann algebra \cite{Aviles:2018jzw} for $D>3$. For $N=3$, the resulting algebra is a Maxwell extension of the post-Newtonian symmetry \eqref{galkmalg} with $\Lambda=0$, first found in \cite{Hansen:2019vqf}. For greater values of $N$ we obtain further post-Newtonian corrections of the electric non-relativistic Maxwell algebra.

\subsubsection*{Carrollian expansions}

Similarly, we can define Carrollian expansions of the Maxwell algebra by using the following alternative $\mathbb{Z}_{2}$-graded subspace decomposition
\begin{align}
V_{0}&=\{ \tilde J_{ab},\tilde P_{a}, \tilde Z_{ab}\}\,, & V_{1}&=\{\tilde G_{a}, \tilde H,\tilde Z_{a}\}
\end{align}
where we have interchanged the generators $H$ and $P_a$ with respect to the Galilean case. This leads to the the resonant non-reduced expanded algebra
 \begin{equation}
\Big(S_{0}^{(\infty)}\times\left\{ \tilde J_{ab},\tilde P_{a}, \tilde Z_{ab}\right\} \Big)\oplus\Big(S_{1}^{(\infty)}\times\left\{ \tilde G_{a}, \tilde H,\tilde Z_{a}\right\} \Big)
 \end{equation}
The expanded in this case take the form
 \begin{equation}
\begin{array}{cc}
J_{ab}^{(m)}=\lambda_{2m}\otimes \tilde  J_{ab}\,,\qquad & B_{a}^{(m)}=\lambda_{2m+1}\otimes \tilde  G_{a}\\
P_{a}^{(m)}=\lambda_{2m}\otimes \tilde  P_{a}\,,\qquad & H^{(m)}=\lambda_{2m+1}\otimes \tilde  H\\
Z_{ab}^{(m)}=\lambda_{2m}\otimes \tilde Z_{ab}\,,\qquad & Z_{a}^{(m)}=\lambda_{2m+1}\otimes \tilde  Z_{a} \,,
\end{array}
 \end{equation}
which leads to the infinite-dimensional Carrollian Maxwell algebra
\begin{multicols}{2}
\begin{subequations}\label{MaxwellinfiniteCar}
\setlength{\abovedisplayskip}{-15pt}
\allowdisplaybreaks
\begin{align}
[  B^{(m)}_{a},  H^{(n)}]&=  P^{(m+n+1)}_{a}\, ,\\
[  B^{(m)}_{a},  P^{(n)}_{b}]&=\delta_{ab}  H^{(m+n)}\, , \\
[  J^{(m)}_{ab},  P^{(n)}_{c}]&= 2\delta_{c[b}  P^{(m+n)}_{a]}\, ,\\
[  B^{(m)}_{a},  B^{(n)}_{b}]&=  J^{(m+n+1)}_{ab}\, ,\\
[  J^{(m)}_{ab},  B^{(n)}_{c}]&=2\delta_{c[b}  B^{(m+n)}_{a]}\, ,\\
[  J^{(m)}_{ab},  J^{(n)}_{cd}]&=4\delta_{[a[c}  J^{(m+n)}_{d]b]}\, ,\\
[  J^{(m)}_{ab},  Z^{(n)}_{cd}]&=4\delta_{[a[c}  Z^{(m+n)}_{d]b]}\, ,\\
[  J^{(m)}_{ab},  Z^{(n)}_{c}]&=2\delta_{c[b}  Z^{(m+n)}_{a]}\, ,\\
[  Z^{(m)}_{ab},  B^{(n)}_{c}]&=2\delta_{c[b}  Z^{(m+n)}_{a]}\, ,\\
[  B^{(m)}_{a},  Z^{(n)}_{b}]&=  Z^{(m+n+1)}_{ab}\, ,\\
[  P^{(m)}_{a},  H^{(n)}]&=-  Z^{(m+n)}_{a}\, ,\\
[  P^{(m)}_{a},  P^{(n)}_{b}]&=   Z^{(m+n)}_{ab}\, .
\end{align}
\end{subequations}
\end{multicols}\noindent
As before, quotients of this algebra by suitable ideals reproduce the $S_E^{(N)}$ expansions, which define Maxwell extensions of the algebras presented in Section \ref{Carroll(A)dSexpansions} for $\Lambda=0$.

\subsection{Non-relativistic brane expansions of (A)dS}
\label{sec:brane}

In order to define non-relativistic $p$-brane expansions of the (A)dS algebra \eqref{AdSalgCommutators} we decompose the relativistic indices 
\begin{align}
A&= ( \alpha,a )\,,& 
\alpha&=0,1,\ldots,p\,,&  a&=p+1,\ldots,D-1\,.
\end{align}
This induces the following decomposition of the generators:
\begin{align}
\tilde J_{AB}&\rightarrow\{\tilde J_{\alpha \beta}, \tilde J_{\alpha a}, \tilde J_{ab}\}\, ,&
\tilde  P_{A}&\rightarrow\{\tilde P_{\alpha}, \tilde P_{a}\}\, .
\end{align}

\subsubsection*{$p$-brane Newton--Hooke expansions}
Non-relativisitic brane expansions of (A)dS of Newton--Hooke type can be defined starting from the following subspace decomposition\,,
\begin{align}
V_{0}&=\{\tilde P_{\alpha}, \tilde J_{\alpha\beta}, \tilde J_{ab}\}\,,&
V_{1}&=\{\tilde P_{a}, \tilde J_{\alpha a}\}\,.\label{eq:splittingStringyGalilei}
\end{align}
Using \eqref{infiniteS0S1} we define the following resonant expansion of (A)dS
 \begin{equation}\label{redrespstringy}
\Big(S_{0}^{(\infty)}\times\{\tilde P_{\alpha}, \tilde J_{\alpha\beta}, \tilde J_{ab}\}
\Big)\oplus\Big(S_{1}^{(\infty)}\times\{\tilde P_{a},\tilde J_{\alpha a}\} \Big)\,.
 \end{equation}
which is spanned by the following expanded generators
\begin{equation}
\begin{aligned} 
H_{\alpha}^{(m)}&=\lambda_{2m}\otimes \tilde  P_{\alpha} \,,&     P_{a}^{(m)}&=\lambda_{2m+1}\otimes \tilde  P_{a} \,,\\
J^{(m)}_{\alpha\beta}&=\lambda_{2m}\otimes \tilde  J_{\alpha\beta}\,,&   B _{\alpha a}^{(m)}&=\lambda_{2m+1}\otimes \tilde  J_{\alpha a} \,,\\
J^{(m)}_{ab}&=\lambda_{2m}\otimes \tilde  J_{ab}\,.\quad &   
\end{aligned}
\end{equation}
The algebra of these generators is given by
\begin{multicols}{2}
\begin{subequations}\label{branegalJG}
\setlength{\abovedisplayskip}{-15pt}
\allowdisplaybreaks
\begin{align}
[J^{(m)}_{\alpha \beta },H^{(n)}_{\gamma}]&= 2\eta_{\gamma[\beta}H^{(m+n)}_{\alpha]}\, , \\
[J^{(m)}_{ab},P^{(n)}_{c}]&= 2\delta_{c[b}P^{(m+n)}_{a]}\, ,\\
[J^{(m)}_{\alpha\beta},J^{(n)}_{\gamma \delta}]&= 4\eta_{[\alpha[\gamma}J^{(m+n)}_{\delta]\beta]}\, , \\
[J^{(m)}_{ab},J^{(n)}_{cd}]&=4\delta_{[a[c}J^{(m+n)}_{d]b]}\, ,\\
[B^{(m)}_{\alpha a},P^{(n)}_{b}]&=\delta_{ab}H^{(m+n+1)}_{\alpha}\, , \\
[B^{(m)}_{\alpha a},H^{(n)}_{\beta}]&=-\eta_{\alpha \beta}P^{(m+n)}_{a}\, ,\\
[  H^{(m)}_{\alpha},   H_\beta^{(n)}]&=-\Lambda   B^{(m+n)}_{\alpha \beta}\, ,\\
[  P^{(m)}_{a},   H_{\alpha}^{(n)}]&=\Lambda   B^{(m+n)}_{\alpha a}\, ,\\
[  P^{(m)}_{a},   P^{(n)}_{b}]&= -\Lambda    J^{(m+n+1)}_{ab}\,,\\
[J^{(m)}_{\alpha \beta},B^{(n)}_{\gamma d}]&=2\eta_{\gamma[\beta}B^{(m+n)}_{\alpha ]d}\, , \\
[J^{(m)}_{ab},B^{(n)}_{\alpha c}]&= 2\delta_{c[b |}B^{(m+n)}_{\alpha | a]}\, ,\\ 
[B^{(m)}_{\alpha a},B^{(n)}_{\beta b}]&=-\eta_{\alpha \beta}J^{(m+n+1)}_{ab}-\delta_{ab}J^{(m+n+1)}_{\alpha \beta}\, .
\end{align}
\end{subequations}
\end{multicols}\noindent

Truncations of this infinite-dimensional algebra by suitable ideals give rise to finite expansions with the semigroup $S_E^{(N)}$ for different values of $N$. The simplest case corresponds to $N=1$, which is the unextended $p$-brane Newton--Hooke algebra in $D$ dimensions \cite{Brugues:2006yd,Andringa:2012uz}. The case $N=2$ gives a higher-dimensional $p$-brane generalisation of the three-dimensional extended stringy Newton-Hooke algebra \cite{Aviles:2019xed}. As in the particle case, higher values of $N$ give rise to post-Newtonian extensions of the brane Newton--Hooke algebra. In the case of vanishing cosmological constant, these reduce to the non-relativistic expansion of the $p$-brane Galilean algebra \cite{Bergshoeff:2018yvt,Bergshoeff:2018vfn,Bergshoeff:2019pij}, which have been studied in \cite{Harmark:2019upf}.

\subsubsection*{$p$-brane Carrollian expansions}

In order to formulate $p$-brane Carrollian expansions, we follow \cite{Barducci:2018wuj} and interchange $\tilde P_\alpha$ and $\tilde P_a$ in the subspace decomposition \eqref{eq:splittingStringyGalilei}, i.e.
\begin{align}
V_{0}&=\{\tilde P_{a}, \tilde J_{\alpha\beta}, \tilde J_{ab}\}\,,&
V_{1}&=\{\tilde P_{\alpha}, \tilde J_{\alpha a}\}\,,
\end{align}
leading to the resonant expanded algebra
 \begin{equation}
\Big(S_{0}^{(\infty)}\times\{\tilde P_{a}, \tilde J_{\alpha\beta}, \tilde J_{ab}\}
\Big)\oplus\Big(S_{1}^{(\infty)}\times\{\tilde P_{\alpha},\tilde J_{\alpha a}\} \Big)\,.
 \end{equation}
and expanded generators
\begin{equation}
\begin{aligned} 
P_{a}^{(m)}&=\lambda_{2m}\otimes \tilde  P_{a}  \,,&     H_{\alpha}^{(m)}&=\lambda_{2m+1}\otimes \tilde  P_{\alpha} \,,\\
J^{(m)}_{\alpha\beta}&=\lambda_{2m}\otimes \tilde  J_{\alpha\beta}\,,&   B _{\alpha a}^{(m)}&=\lambda_{2m+1}\otimes \tilde  J_{\alpha a} \,,\\
J^{(m)}_{ab}&=\lambda_{2m}\otimes \tilde  J_{ab}\,.\quad &   
\end{aligned}
\end{equation}
This yields the following infinite-dimensional Carrollian (A)dS algebra
\begin{multicols}{2}
\begin{subequations}\label{braneCarrollJG}
\setlength{\abovedisplayskip}{-15pt}
\allowdisplaybreaks
\begin{align}
[J^{(m)}_{\alpha \beta },H^{(n)}_{\gamma}]&= 2\eta_{\gamma[\beta}H^{(m+n)}_{\alpha]}\, , \\
[J^{(m)}_{ab},P^{(n)}_{c}]&= 2\delta_{c[b}P^{(m+n)}_{a]}\, ,\\
[J^{(m)}_{\alpha\beta},J^{(n)}_{\gamma \delta}]&= 4\eta_{[\alpha[\gamma}J^{(m+n)}_{\delta]\beta]}\, , \\
[J^{(m)}_{ab},J^{(n)}_{cd}]&=4\delta_{[a[c}J^{(m+n)}_{d]b]}\, ,\\
[B^{(m)}_{\alpha a},P^{(n)}_{b}]&=\delta_{ab}H^{(m+n)}_{\alpha}\, , \\
[B^{(m)}_{\alpha a},H^{(n)}_{\beta}]&=-\eta_{\alpha \beta}P^{(m+n+1)}_{a}\, ,\\
[  H^{(m)}_{\alpha},   H_\beta^{(n)}]&=-\Lambda   B^{(m+n+1)}_{\alpha \beta}\, ,\\
[  P^{(m)}_{a},   H_{\alpha}^{(n)}]&=\Lambda   B^{(m+n)}_{\alpha a}\, ,\\
[  P^{(m)}_{a},   P^{(n)}_{b}]&= -\Lambda    J^{(m+n)}_{ab}\,,\\
[J^{(m)}_{\alpha \beta},B^{(n)}_{\gamma d}]&=2\eta_{\gamma[\beta}B^{(m+n)}_{\alpha ]d}\, , \\
[J^{(m)}_{ab},B^{(n)}_{\alpha c}]&= 2\delta_{c[b |}B^{(m+n)}_{\alpha | a]}\, ,\\ 
[B^{(m)}_{\alpha a},B^{(n)}_{\beta b}]&=-\eta_{\alpha \beta}J^{(m+n+1)}_{ab}-\delta_{ab}J^{(m+n+1)}_{\alpha \beta}\, .
\end{align}
\end{subequations}
\end{multicols}\noindent
The $N=1$ truncation of this infinite-dimensional symmetry corresponds to the $p$-brane Carroll (A)dS algebra in $D$ dimensions \cite{Clark:2016qbj}, while $S_E^{(N)}$ expansions for greater values of $N$ define post-Carrollian (A)dS $p$-brane symmetries. See also~\cite{Roychowdhury:2019aoi} for further discussions of Carrollian branes.

\providecommand{\href}[2]{#2}\begingroup\raggedright\endgroup

%\bibliographystyle{utphys}
%\bibliography{references}

\end{document}